\documentclass [12pt] {report}
\usepackage{amssymb}
\newcommand {\D}[2] {\displaystyle\frac{\partial{#1}}{\partial{#2}}}

\newcommand {\Dd}[3] {\displaystyle\frac{\partial^2{#1}}{\partial{#2}\partial{#3}}}

\newcommand {\ga} {\gamma}
\newcommand {\la} {\lambda}
\newcommand {\La} {\Lambda}
\newcommand {\si} {\sigma}
\newcommand {\Si} {\Sigma}
\newcommand {\de} {\delta}
\newcommand {\prtl} {\partial}
\newcommand {\fr} {\displaystyle\frac}
\newcommand {\wt} {\widetilde}
\newcommand {\be} {\begin{equation}}
\newcommand {\ee} {\end{equation}}
\newcommand {\ba} {\begin{array}}
\newcommand {\ea} {\end{array}}
\newcommand {\bp} {\begin{picture}}
\newcommand {\ep} {\end{picture}}
\newcommand {\bc} {\begin{center}}
\newcommand {\ec} {\end{center}}
\newcommand {\bt} {\begin{tabular}}
\newcommand {\et} {\end{tabular}}
\newcommand {\lf} {\left}
\newcommand {\rg} {\right}

\newcommand {\cF} {{\cal F}}

\newcommand {\cR} {{\cal R}}
\newcommand {\cS} {{\cal S}}

\newcommand {\ses} {\medskip}

\newcommand {\e} {\mathop{\rm e}\nolimits}

\newcommand {\g}  {\stackrel{g\to -g}{\Longleftrightarrow}}

\newcommand {\cP} {{\cal P}}
\newcommand {\bibit} {\bibitem}
\newcommand {\nin} {\noindent}

\newcommand {\Ga} {\Gamma}
\newcommand {\De} {\Delta}

\newcommand {\Rho} {\mbox{\large$\rho$}}

\newcommand {\ka} {\varkappa}

\usepackage{epsfig, latexsym,graphicx}
 \usepackage{amsmath}

\def\2#1#2#3{{#1}_{#2}\hspace{0pt}^{#3}}
\def\3#1#2#3#4{{#1}_{#2}\hspace{0pt}^{#3}\hspace{0pt}_{#4}}
\newcounter{sctn}
\def\sec#1.#2\par{\setcounter{sctn}{#1}\setcounter{equation}{0}
                  \noindent{\bf\boldmath#1.#2}\bigskip\par}

\begin {document}

\begin {titlepage}

\vspace{0.1in}

\begin{center}

{\Large \bf Finsleroid  Corrects     Pressure  and Energy   of}\\

\ses

\ses

{\Large \bf  Universe. Respective  Cosmological Equations }

\end{center}

\vspace{0.3in}

\begin{center}

\vspace{.15in} {\large G.S. Asanov\\} \vspace{.25in}
{\it Division of Theoretical Physics, Moscow State University\\
119992 Moscow, Russia\\
{\rm (}e-mail: asanov@newmail.ru{\rm )}} \vspace{.05in}

\end{center}

\begin{abstract}

The Hubble constant proves to be  the
pseudo-Finsleroid--Landsberg factor. The  covariantly conserved
pseudo-Finsleroid--gravitational tensor is  explicitly  found
after evaluating the  respective Finsleroid--case curvature tensor
and required contractions
 in attentive  way.
The  equations arisen involve one  parameter $g$ of extension
which measures the Finslerian deviation of the curvature of the
indicatrix of unit vectors. The vector field $b^i(x)$ of the axes
of the pseudo--Finsleroids is naturally identified to the  field
of average velocity vectors of matter of the universe. The
consistent (and unique) continuation of the Robertson--Walker
metric, and hence the Friedmann metrics,
 in the Finslerian domain with
respect to the    parameter $g$ is arisen. The cosmological
pressure and energy density prove to be  linear functions of
$g^2$, so that the presence of the negative pressure seems to be
not necessary to get the agreement with  the observed negative
nature of deceleration parameter.
We clarify the explicit structure of all the involved tensorial objects.

\ses

\nin {\it  Key words}: Cosmological equations, energy density,
pressure, Finsleroid geometry, relativity.

\end{abstract}

\end{titlepage}

\vskip 1cm

\setcounter{sctn}{1}
\setcounter{equation}{0}

\nin
{\bf 1.  Introduction}

\bigskip

To pattern the physics of the  universe in terms of Finslerian as well as
Riemannian equations, we are to comply with the observation
evidence that to a great extent the universe is homogeneous and
isotropic. It is well-known that the evidence is perfectly
 matched to the Friedmann-Robertson-Walker metrics formulated against the (pseudo-)Riemannian geometrical background
(see  [1,2]).

It is amazing, however,  that the metrics
can straightforwardly be extended to the (pseudo-)Finsleroid---geometry
framework, strictly preserving both the homogeneity and isotropy.
The Finsleroid charge $g$ is the parameter of the extension.

Having had the far--developed Finsleroid geometry at hands,
we can follow a new and lifted geometric route to search for due extensions of equations of
 the cosmological theory.
What might happen if the cosmological value of the parameter $g$ is essential,  what if   $g=2$?
The pending answer in general is:  the cosmological ``puzzles''
 may occur to be not manifestations of some new and exotic matter constituents,
but merely implications  of the Finsleroid--extensions of field  equations,  --- at least contributions from such
extensions are to be accounted for.

The nearest and   simple consequence of the Finsleroid--case  cosmological
 equations is that they may allow for the presence
of the negative deceleration parameter without assuming the negative pressure.
The systematic description of the observed phenomenon of  the accelerated expansion  of the universe  can be found
in  the  modern  review [2].

{

In the Finsleroid--geometry context,
the phenomenon may merely communicate us  of the presence of the pseudo--Finsleroid charge $g$.
Indeed, whenever the  three--dimensional space is flat
 the conventional  cosmology theory proposes  the pressure
$$
p_{\text{Friedmann}}=-\lf(1-2q_{\text{cosm}} \rg)H^2
$$
(normalized by $8\pi G$),
where $H$ is the Hubble constant; by  $q_{\text{cosm}}$
we denote the deceleration parameter.
The Finsleroid--case pressure
$$
p=-\lf(1-2q_{\text{cosm}}+\fr{g^2}4+g^2 q_{\text{cosm}} \rg)H^2
$$
(see (1.54))    provides us with the extension  which involves the square $g^2$
of the pseudo--Finsleroid charge.
One  may conclude that
$$
(2-g^2)  q_{\text{cosm}}  =     1+\fr{g^2}4, \qquad \text{whenever} \quad p=0.
$$
If we tentatively put here
$q_{\text{cosm}}  =-1$,
 we obtain $|g|=2$. Thus,  the negative value of
the deceleration parameter $q_{\text{cosm}}$ may well be compatible with the zero pressure
if the  Finsleroid charge $g$ is not postulated to be  zero.

The energy density of the universe also obtains corrections with respect to the square $g^2$
(see (1.56)).

{

The principal distinction of the Finsleroid geometry (generally, of the Finsler geometry)
from the Riemannian geometry may be seen in the
circumstance that the associated Finslerian metric tensor  $g_{mn}(x,y)$
becomes being dependent on the pair $(x,y)$, --- the so--called {\it line element}, ---
and hence on  the tangent vectors $y$ (which are supported by $x$).
Such a dependence  is succeeded to the curvature tensor  $R_n{}^i{}_{mj}(x,y)$ and, after that,
to  the  concomitant tensors and vectors
$$
\rho_{ij}=  \rho_{ij}(x,y),  \qquad     \rho^i=  \rho^i(x,y)
$$
which arise upon extension of their relativistic gravitational precursors.

On the other hand,
the deviation equation (1.30), as well as the
geodesic equation (1.29),
obviously belongs to the {\it Category of Observable Concepts}.
 In these equations, the vector $y$ has the meaning of the
velocity of a test particle
(which does not contribute to the local gravitational field).
Therefore, the variable $y$
can be given  the interpretation of the vector of a local observer located at the point $x$ from which the vector $y$
is issued.
Following this motivation, we should interpret the dependence of $\{\rho_{ij},\,\rho^i\}$
 on $y$ to be dependence
on the velocity vector of the motion state of a local test--observer.

At the same time, in the cosmological pattern we have globally the {\it cosmological reference frame}
produced by   the  field $b^i_{\text{cosm}}(x)$ of average velocity vectors of matter of the universe.
Geometrically, the pseudo--Finsleroid is a rotund (axial--symmetric) body. Denote by
 $b^i_{\text{pseudo--Finsleroid}}(x)$ the vector field
which is such that at any given point $x$
the  vector  $b^i_{\text{pseudo--Finsleroid}}(x)$ assigns the axis of the pseudo--Finsleroid.
It is quite natural to apply the identification
$$
b^i_{\text{cosm}}(x)= b^i_{\text{pseudo--Finsleroid}}(x)\equiv b^i(x)
$$
when starting to initiate the Finsleroid--extension of cosmological theory.

\ses\ses

Accordingly, we shall comply with the following principle.

\ses\ses

{\large The  Osculation--Correspondence Principle}.
 The $x$--dependent extensions, which are functions of the points $x$ of the underlying
 manifold,  stem from their
$(x,y)$--dependent counterparts  through the osculation along the fundamental
vector field $b^h(x)$:
$$
y^h ~ \to ~ b^h(x).
$$
The fields thus appeared are called {\it osculating} to their
$(x,y)$--dependent counterparts along the vector field $b^h(x)$.

\ses\ses

{

In particular, the field
$
\rho^i_{\{b\}}(x)~:=\rho^i_{|_{y^h=b^h}}
$
(see (1.39))
is
osculating to the field $\rho^i(x,y)$.
In this equality, the left--hand part is osculating to the right--hand part
along the vector field $b^h(x)$.
The equality
$
g_{ij}(x,b^h(x))= a_{ij}(x)
$
(see (A.29) in Appendix A) means:  the associated Riemannian metric tensor
$a_{ij}(x)$ is osculating to the Finsleroid metric tensor
$g_{ij}(x,y)$ along the vector field $b^h(x)$.
The following important equality holds:
$$
g_{ij}(x,y)_{|_{g=0}}=g_{ij}(x,b^h(x)).
$$

We can naturally interpret the osculating--cosmological  fields as the
{\it hydrodynamic} fields, that is, the fields which describe the universe at the hydrodynamic level of
comprehension.
In this vein, the associated Riemannian metric tensor
$a_{ij}(x)$ is the hydrodynamic--level metric tensor of the universe.
The pressure $p(x)$ used above is arisen upon
the osculating--hydrodynamic method from the general $(x,y)$--dependent field (see  (C.67) in Appendix C),
--- the same novelty should be said about the energy density $\Rho(x)$ (see (C.66) in Appendix C).

It is a striking fact   that the form (1.40) of the continuity equation
is not affected by the Finsleroid extension (see also Proposition 1.4).


When pondering upon possibility to extend the cosmology theory in a Finsler way,
the robust care should exercised that
the general--status Finsler geometry (see [3-10])
does not provides us with a direct
possibility  to extend the gravitational field equations.
 This particular circumstance is dramatically at variance with what we have
in the framework of the Riemannian geometry.
 Indeed, in the latter geometry,  the Riemannian curvature tensor, --- to be denoted by $a_i{}^j{}_{mn}$, --- produces
uniquely the Riemannian Ricci tensor $a_i{}^j{}_{jn}$ on contraction. When one uses the last tensor
 to construct the gravitational field tensor
$E^{\text{\{Riem\}}}{}_{in}=
a_i{}^j{}_{jn} - \frac12 a_{in}a^{mk}{}_{mk}$,
one may get the  implication from appropriate identities obeyed by
the curvature tensor   $a_i{}^j{}_{mn}$  that the tensor $E^{\text{\{Riem\}}}{}_{in}$ is covariantly conserved:
 $\nabla_i E^{\text{\{Riem\}}}{}_{n}{}^i=0$, where    $\nabla_i$ stands for the Riemannian covariant derivative.
The  general--status  Finsler geometry  does not provide us with  a similar lucky implication.
In the Riemannian geometry, the equality       $a_i{}^j{}_{jn}=a^j{}_{inj}$   holds,
but in the Finsler geometry we are faced with
$R_i{}^j{}_{jn} \ne R^j{}_{inj}$.
Moreover, neither the contraction $R_i{}^j{}_{jn}$
nor the contraction $R^j{}_{inj}$ is symmetric with respect to the indices $i,n$, so that
 the Ricci tensor of the Riemannian geometry cannot be unambiguously extended to the domain
of general Finsler geometry.

{

However, in the Finsleroid--Landsberg space applied in the present paper
we have at our disposal the tensor  $\rho_{ij}$ (defined by (1.31))
which is  covariantly--conserved (see (1.32)) and, therefore, is attractive to
 be used as a required extension of the gravitational
tensor $E^{\text{\{Riem\}}}{}_{in}$.
 The entailed field  $\rho^i=\rho^i{}_j y^j$ (see (1.33) and (1.34)) can naturally be treated as the current
 of energy.

The Finsleroid--based analysis involves
 a long chain of calculations, which are simple but sometimes tedious.
The very method, however,  can be explained in simple words.

{

 Namely,
let us start with a positive--definite Riemannian space $R_N$ of the dimension $N$, referred to local coordinate set
$\{x^i\}$. Denote by $a_{mn}$ the Riemannian metric tensor of the space $R_N$
and  by $\nabla_n$ the associated Riemannian covariant derivative.
Assume that a covariant vector field $b_i(x)$ be given on the space  $R_N$. Set forth the unit length condition
\be
a_{mn}b^mb^n=1,
\ee
where
$b^m=a^{mn}b_n$ is the associated contravariant vector field.
We want to specify the covariant derivative $\nabla_nb_i$  in a manner attractive to apply
in cosmological consideration. Let us choose the possibility
\be
\nabla_nb_m=k(a_{mn}-b_mb_n),
\ee
where
\be
 k=k(x)
\ee
is a scalar.
From  (1.2) we conclude that the 1-form $b=b_i(x)dx^i$ must be closed:
\be
\partial_ib_j-\partial_ib_i=0,
\ee
where $ \partial_i=\partial/\partial x^i$. From (1.1) and (1.2) it follows that the field $b^i(x)$ is geodesic:
\be
b^n\nabla_nb_m=0.
\ee
It is convenient to introduce the {\it projection tensor}
\be
r_{mn}=a_{mn}-b_mb_n,
\ee
which fulfills the identities
\be
b^mr_{mn}=b^nr_{mn}=0.
\ee
In terms of the tensor (1.6) the condition (1.2) reads merely
\be
\nabla_nb_m=kr_{mn}.
\ee

We shall also use the notation
\be
 k_n=\partial_n k
\ee
and
\be
(bk)=b^nk_n,
\ee
together with
\be
\wt k_n=\partial_n k+k^2b_n
\ee
and
\be
(b\wt k)=b^n \wt k_n.
\ee

{

If we  introduce in  the space $\cR_N$
 a  $b$--{\it geodesic coordinate set}
$\{z^A\},\, A=\{a,N\}, a,b,c,d=1,...,N-1,$
  arisen upon an admissible coordinate transformation $x^i=x^i(z^A)$
such that the vector field $b^i(x)$
occurs featuring tangent to the $z^N$-coordinate line,
then it is easy to see that with respect to the coordinates
$\{z^A\}$
\be
b^N=1, \qquad    b_a=0, \qquad  a_{NN}=1,  \qquad  a_{Na}=0, \qquad  r_{NN}=0, \qquad  r_{Na}=0,
\ee
and the squared Riemannian metric $(ds)^2=a_{mn}dx^mdx^n$
is transformed to the sum
\be
(ds)^2=(dz^N)^2+r_{ab}(z)dz^adz^b.
\ee
Evoke the {\it factorization condition}
\be
r_{ab}(z^A)=\lf(\phi(z^A)\rg)^2 p_{ab}(z^c),
\ee
so that the tensor $p_{ab}(z^c)$ is assumed to be independent of the coordinate $z^N$.
By rewriting the condition (1.8) it terms of the coordinate $\{z^A\}$ it is easy to see that
the factor $k$ entering (1.8) is obtainable through
\be
k=\fr1{\phi}\D{\phi}{z^N}.
\ee

{

Let us set forth also the condition
$$
\D{\phi(z^A)}{z^a}=0,
$$
that is,
\be
\phi=\phi(z^N).
\ee
Converting this condition to the  initial coordinates $x^i$ yields
\be
k_n=(bk)b_n,
\ee
which entails the representation
\be
\wt k_n=(b\wt k)b_n
\ee
for the vector (1.11).
With  postulating (1.15), it is convenient to use the notation
\be
\dot k=\frac {dk}   {dz^N},
\ee
getting
\be
\fr{\dot\phi}{\phi}=k,
\qquad
\dot k+k^2= \fr1{\phi}   \ddot \phi,
\ee
together with
\be
(bk)=\dot k
\ee
and
\be
 (b\wt k)=\dot k+k^2.
\ee

Thus the following proposition is a truth.

\ses\ses

{\large Proposition 1.1.} {\it  When in addition to
$$
a_{mn}b^mb^n=1, \qquad  \nabla_nb_m=kr_{mn},
\qquad   k_n=(b k)b_n,
$$
the factorization condition}
 (1.15)
 {\it    is applied,  the Riemannian space  $R_N$ is of the  warped type

}
\be
(ds)^2=(dz^N)^2+    \lf(\phi(z^N)\rg)^2 p_{ab}(z^c)  dz^adz^b.
\ee

\ses\ses

{\large Definition.}
Under the conditions of Proposition 1.1 we say that we deal with the {\it special case}
of the Riemannian space $\cR_N$.

\ses\ses

{\large Note.}
Obviously, the conditions (1.17) and (1.18) are equivalent, so that
the special case can equivalently be characterized by the condition (1.18) instead of the condition (1.17).

\ses\ses

{

To lift the consideration to the Finsler--geometry level with the aim of specifying an appropriate
Finslerian metric function $K(x,y)$,
 where $y$ are tangent vectors,    we  assume
that the sought  function
$K(x,y)$ be of the functional dependence
\be
K(x,y) =\Phi^{PD}\Bigl(g(x), b_i(x), a_{ij}(x),y\Bigr),
\ee
where $g(x)$ is a scalar playing the role of the {\it Finslerian parameter} of extension
and $\Phi$ is a scalar which smoothness is to be of at least class  $C^4$ (if  the class
$C^{\infty}$ is not realized).

\ses \ses

FINSLEROID  PRINCIPLE  OF  EXTENSION.
In each tangent space, the              indicatrix
(=surface of unit-length vectors defined by the function  $K$)
 is to be of constant curvature.
The indicatrix is rotund with the vector $b^i=a^{ij}b_j$ being the axial symmetry. The indicatrix is
closed and everywhere regular and the resultant Finsler space is positive--definite.

 \ses\ses

Under these conditions, we call the unit ball bounded by the indicatrix
the {\it Finsleroid}.

For the curvature $\cR$ of the indicatrix, the value
\be
 \cR_{\text{Finsleroid  Indicatrix} }= 1-\fr{g^2}4
\ee
is obtained.
Thereby the metric function $K$ together with the entailed  $\cF\cF^{PD}_g $--space
(the meaning of the upperscripts $PD$ reads ``positive--definite'')
is completely and uniquely determined (see the formulas (A.1)--(A.16) in Appendix A).
Notice that the curvature value (1.26) is independent of neither the tensor $a_{ij}(x)$
nor the field   $ b_i(x)$.

The continuation parameter $g$ is dimensionless, being of  {\it pure--geometrical meaning.}

\ses\ses

CRUCIAL  DEFINITION.
The $\cF\cF^{PD}_g $--space subjected to the conditions
$\nabla_nb_m=kr_{mn}$
and
\be
g=const
\ee
is called        the {\it Finsleroid--Landsberg space}.
The scalar $k$ is called the {\it Finsleroid--Landsberg factor}.

\ses

{

In any Finsler space (see the books [3-5]), the geodesic equations are defined  in the unique way
by the help of the so--called spray coefficients
\be
G^i~:=\ga^i{}_{mn}(x,y)y^my^n
\ee
with
 $\ga^i{}_{mn}$
standing for  the Finslerian Christoffel symbols constructed from the Finsleroid--Finsler  metric function $K$.
The geodesic  equations read
\be
\fr{d^2 x^i}{d\si^2} + G^i(x,y)=0,  \qquad y^h=\fr{dx^h}{d \si},
\ee
where $d\si=\sqrt{|g_{mn}(x,dx)dx^mdx^n|}$
is the Finslerian arc--length.

The geodesic deviation equation reads
\be
\fr{\de^2\eta^i}{d\si^2} + R_n{}^i(x,y) \eta^n=0, \qquad y^h=\fr{dx^h}{d \si}.
\ee
The deviation tensor $ R_n{}^i(x,y)$  thus appeared entails the full curvature tensor
$R_n{}^i{}_{mj}(x,y)$
 upon applying appropriate differentiation with respect to the variable $y$
 (see (A.59)--(A.61) in Appendix A).

{

The known fact (indicated in [11-14])
is that
{\it in the Finsleroid--Landsberg space
  the tensor
\be
\rho_{ij}~:=\fr12(R_i{}^m{}_{mj}+R^m{}_{ijm})-\fr12g_{ij}R^{mn}{}_{nm}
\ee
is covariantly conserved:
\be
\rho^i{}_{j|i} \equiv 0.
\ee
}

\ses

The covariant derivative $|i$ is constructed according to the known Finslerian rule (see  (A.68)--(A.70) in Appendix A).

Let us construct by means of contraction  the {\it produced current}
\be
\rho^i~:=\rho^i{}_j y^j.
\ee
Since $y^j_{|n}=0$ in any Finsler space,
from (1.32) it issues that the following proposition is valid.

\ses\ses

{\large Proposition 1.2.} {\it In the Finsleroid--Landsberg space
 the  current $\rho^i$
is  covariantly conserved:
\be
\rho^i_{|i} \equiv 0.
\ee

}

\ses



\ses\ses

{\large Definition.}
The $\cF\cF^{PD}_g $--space is said to be of
the {\it isotropic case} if the conditions claimed in Proposition 1.1 are fulfilled and
the tensor $p_{ab}$ entered the metric (1.24)
is such that  the Riemannian curvature tensor evaluated from the tensor corresponds to the
constant--curvature case (that is, the representations (C.32)--(C.34) of Appendix C hold).

The current can be  expanded  with respect to the basis $\{ b^i, y^i \}$
\be
\rho^i=  c_1(x,y)b^i+c_2(x,y)y^i
\ee
(the coefficients $c_1,c_2$ can be found from the formula (C.61) of Appendix C)
or, alternatively,
   with respect to the basis $ \{y^i,A^i\}$
\be
\rho^i= n_1(x,y)y^i+n_2(x,y)A^i
\ee
(the coefficients $n_1,n_2$ can be found from the formula (C.58) of Appendix C), where
$A^i$ is the contraction of the Cartan tensor of the Finsler geometry
(see (A.34)--(A.38) in Appendix A).

{

\ses

In the dimension
$$N=4$$
from  (C.68) we may conclude that the quantity
\be
p~:=-\cP
\ee
equals
\be
p=-\Biggl[\ka +\lf(1-\fr{g^2}4\rg)k^2 +(2+g^2)(b\wt k)\Biggr]
\ee
(the equality (C.44) has been used),
and
with the definition
\be
\rho^i_{\{b\}}(x)~:=\rho^i_{|_{y^h=b^h}}
\ee
of the Finsleroid--extension of the current
from (C.70) and (C.71) we obtain
\be
 \nabla_i \rho^i_{\{b\}} =  -  3k p.
\ee

{

In the space--time context proper, we take  the dimension $N=4$ and apply the pseudo-Finsleroid
space.
A close inspection shows that we may straightforwardly perform the convertation
\be
K\to F
\ee
of  the positive--definite Finsleroid metric function $K$ into
the relativistic--case pseudo-Finsleroid metric function, to be denoted by $F$,
by applying the formal change
\be
g\to ig
\ee
(where $i$ is the imaginary unity),
taking the associated metric tensor $a_{mn}(x)$ to be of the pseudo--Riemannian type
\be
a_{mn}(x)=b_m(x)b_n(x)-r_{mn}(x)
\ee
with the  condition of positive--definiteness
\be
r_{mn}y^my^n > 0  \quad \text{and} \quad \text{rank}(r_{mn})=N-1
\ee
(we always assume $y\ne 0$).
The covariant vector field $b_i(x)$ is to be  taken time--like and  the unit condition
(1.1) persists.
The indicatrix curvature  value (1.26) is replaced by
\be
 \cR_{\text{pseudo-Finsleroid  Indicatrix} }=-\lf(1+\fr{g^2}4  \rg).
\ee
The respective scalar $g(x)$ will be called the {\it pseudo-Finsleroid charge}.
Similarly to (1.25),
the pseudo--Finsleroid metric function $F$ is  of the functional structure
\ses
\be
F(x,y) =\Phi^{SR} \Bigl(g(x), b_i(x), a_{ij}(x),y\Bigr).
\ee
We denote  the resultant space by      the $\cF\cF^{SR}_g $--{\it space}
(with the upperscripts $SR$ meaning ``special--relativistic'').
The Landsberg condition  preserves its form (1.8).
With the Landsberg condition assumed to hold,
we call the $\cF\cF^{SR}_g $--space        the {\it pseudo-Finsleroid--Landsberg space}
and call the scalar $k$  the {\it  pseudo-Finsleroid--Landsberg factor}.

In the cosmological application,
the vector field $b^i(x)=a^{ij}(x)b_j(x)$ of the pseudo--Finsleroid axes is naturally identified with
the  field of average velocity vectors of matter of the universe, thereby
{\it the field   $b^i(x)$ represents   the  co--moving reference frame of the universe.}
Under these conditions, we call the covariantly conserved tensor
 $\rho_{ij}$ given by (1.32) the {\it Finsleroid--gravitational tensor}
and  the vector field $\rho^i$ obtained in the way (1.33)
 the    {\it current of energy of the universe}.

{

\ses

In such a relativistic--cosmological  physical case, we are to make in the warped--metric representation (1.24)
the change
\be
z^N\to z^0=t,  \qquad
\phi\to L,
\ee
with
$t$ meaning the {\it cosmological time},
and with
\be
L=L(t)
\ee
getting the sense of  the   {\it cosmological scale factor}.
Accordingly, the equalities (1.21) change to
\be
\fr{\dot L}{L}=k, \qquad    \dot k+k^2= \fr1L   \ddot L,
\ee
so that the  $k$  acquires the  meaning of the  {\it Hubble constant.}
Also,
\be
(b\wt k)= \dot k+k^2=-\fr{q_{\text{cosm}}}{k^2},
\ee
where $q_{\text{cosm}}$ makes sense of  the cosmological {\it deceleration parameter}.
In the isotropic case  we have
\be
\ka=\fr1{L^2}\ka_1,  \qquad \ka_1=-1, 0, 1
\ee
(see (C.32)--(C.34) in Appendix C).

Under these circumstances, the metric  (1.24) changes to read
\be
(ds)^2=(dt)^2-    \bigl(L(t)\bigr)^2 p_{ab}(z^c)  dz^adz^b.
\ee

\ses

Thus we are arriving at

\ses\ses

{\large Proposition 1.3.} {\it  The  Finsleroid--Landsberg space thus arisen
is the }({\it unique}) {\it continuation of the Robertson-Walker  metric in the  Finslerian domain.
It is the Hubble constant $H$ that  plays the role of   the pseudo-Finsleroid---Landsberg factor $k$:
}
\be
H=k.
\ee

\ses\ses

Since
the cosmological  Friedmann metrics are particular cases of the Robertson-Walker metric, we also
obtain due pseudo-Finsleroid--Landsberg  continuations of the  metrics.

{

In this way we arrive at  the {\it Finsleroid  hydrodynamic pressure}
\be
p=-\lf(1-2q_{\text{cosm}}+\fr{g^2}4+g^2 q_{\text{cosm}} \rg)H^2 -\ka
\ee
(from (1.38)),
and the {\it Finsleroid  energy current}
\be
 \Rho^i= \Rho b^i
 \ee
\ses
together with  the {\it Finsleroid  energy density}
\be
 \Rho= 3\lf(1+\fr{g^2}4\rg)H^2 + 3\ka
\ee
(the formulas (C,64) and (C.66) of Appendix C have been used).
From (1.40) we get the continuity equation
\be
 \nabla_i  \Rho^i =  -  3H p.
\ee
Applying  (A.109) and  noting
$$
b^i\partial_i \Rho=  \fr{d \Rho}{dt}
$$
together with
\be
\nabla_i b^i=3k
\ee
(consider the Landsberg--case condition (1.2)  at $N=4$),
we may write the equation (1.57) in the form
\be
\fr{d\Rho}{dt} +3 H \Rho    =
 -  3 Hp.
\ee
We can also write this as
\be
\fr{d\Bigl( \Rho L^3\Bigl)}{dL}= -3 p  L^2,
\ee
which is precisely of the form characteristic of the respective Friedmann--case equation.
The Finsleroid charge $g$ does not enter the last equation.

Thus we are entitled to formulate

\ses\ses

{\large Proposition 1.4.} {\it The Finsleroid--Landsberg extension of the Friedmann metrics does not
change the conventional equation which describes the energy density of the universe.
When the pressure vanishes, that is
 $p=0$, we obtain from (1.60) the well-known Friedmann--case law
 \be
 \Rho ~ \sim ~ \fr1{ L^3}.
\ee

}

\ses\ses

{

\ses\ses
{\large NOTE}. In the conventional  cosmology theory, the pressure is
\be
p_{\text{Friedmann}}=-\lf(1-2q_{\text{cosm}} \rg)H^2-\ka.
\ee
The Finsleroid--case pressure $p$ given by (1.54)  extends this version (1.62)
by the terms which involve the square of the pseudo--Finsleroid charge, namely $g^2$.

{\it Whenever the three--dimensional space is flat:}
\be
\ka=0,
\ee
 from (1.56)
we may
conclude the validity of  the law
\be
\fr{\text {Finsleroid-density of energy }}
{\text {Friedmann-density of energy }}
=
1+\fr{g^2}4
~ ~   \equiv
- \cR_{\text{pseudo-Finsleroid  Indicatrix} }
\ee
and the pressure (1.54) becomes such that
\be
\fr{\text {Finsleroid pressure}}
{\text {Friedmann pressure }}
=\fr{1-2q_{\text{cosm}}+\fr{g^2}4+g^2 q_{\text{cosm}} }
{1-2q_{\text{cosm}} }
\ee

{

When the  special--case conditions  (C.1)--(C.12) and the isotropy conditions (C.32)--(C.34),  --
set forth in Appendix C,  --  are fulfilled,
the tensor $\rho^{ik}$ can be evaluated in the form of the expansion
with respect to the basis
 $\{g^{ik}, A^i, y^i\}$
\be
\rho^{ik}= \fr {M_7 } {2K^2} g^{ik}
+\fr {M_8}   {2K^2}\fr1{A_lA^l}A^iA^k
  -\fr{ 2q}{NKg}\fr 1{K^2}       \Bigl(Y_{\{y\}}y^kA^i + Z_{\{e\}}A^ky^i \Bigr)
+\fr {M_{10} }   {2K^4} y^iy^k
\ee
or  with respect to the basis
 $\{a^{ik}, b^i, y^i\}$
\be
\rho^{ik}= E_1 a^{ik}
+E_2b^ib^k+E_{3Y}b^iy^k+E_{3Z}b^ky^i + E_4y^iy^k.
\ee
The involved coefficients $E_1,E_2,E_{3Y},E_{3Z},E_4$
depend not only on point $x$ but also on tangent vectors $y$ supported by the $x$.
The explicit form of  the coefficients
can be found in Appendix D.
Unless $g=0$, the non--symmetry
\be
E_{3Y}  \ne E_{3Z},   \qquad
Y_{\{y\}}   \ne   Z_{\{e\}}
\ee
entailing
\be
\rho_{ij} \ne \rho_{ji}
\ee
        takes place.
The skew part
$$\rho^{ik}-\rho^{ik}$$
is explicitly evaluated in Appendix E.

The coefficients $E_{3Y}$,  $E_{3Z}$, and $E_4$  are proportional to the Finsleroid parameter $g$:
\be
E_{3Y}=gT_{3Y}, \qquad  E_{3Z}=gT_{3Z}, \qquad  E_{4}=gT_{4}
\ee
(see (D.40), (D.42), and (D.45)).
They disappear, therefore,   in the Riemannian limit:
\be
(E_{3Y})_{|g=0}=(E_{3Z})_{|g=0}=(E_{4})_{|g=0}=0.
\ee

We also have
\be
(E_1)_{|g=0}=\cP
\ee
(from (D.35) and (D.23))
and
\be
(E_2)_{|g=0}=-(N-2)\bigl(\xi+ (b\wt k)   \bigr), \qquad \xi=-k^2- \ka, \quad (b\wt k)=\dot k+k^2
\ee
(from (D.36) and (D.27)).

The term $\La g_{ij}$ with the cosmological constant $\La$
can also be added to the tensor $\rho_{ij}$
not destroying the validity of the law  of the covariant--divergence conservation.

{

\vskip 2cm

\ses

\ses

\ses

\setcounter{equation}{0}

\nin
{\bf \large  Appendix A: ~ Initial knowledge of   the $\cF\cF^{PD}_g $--objects}

\ses\ses

Let $\cR_N=\{M, a_{mn}\}$ be a Riemannian space, where $M$ is an $N$--dimensional manifold
and $a_{mn}$ is a positive--definite Riemannian metric tensor.
 We introduce on  $M$  a scalar field $g=g(x)$   subject to ranging
\be
-2<g(x)<2,
\ee
and apply  the convenient notation
\be
h=\sqrt{1-\fr14g^2}, \qquad
G= \fr gh.
\ee
The {\it  characteristic  quadratic form}
\be
B(x,y) :=b^2+gqb+q^2
\equiv\fr12\Bigl[(b+g_+q)^2+(b+g_-q)^2\Bigr]>0,
\ee
where $ g_+=\frac12g+h$ and $ g_-=\frac12g-h$,
is of the negative discriminant
\be
D_{\{B\}}=-4h^2<0
\ee
and, therefore, is positively definite.
In the limit $g\to 0$,
the definition (A.3) degenerates to the
 quadratic form of the input Riemannian metric tensor:
\be
B|_{_{g=0}}=a_{mn}y^my^n.
\ee

{

\ses

\ses

 {\large  Definition}.
  The scalar function $K(x,y)$ given by the formulas
\be
K(x,y)=
\sqrt{B(x,y)}\,J(x,y)
\ee
and
\be
J(x,y)=\e^{-\frac12G(x)f(x,y)},
\ee
where
\be
f=
-\arctan \fr G2+\arctan\fr{L}{hb},
\qquad  {\rm if}  \quad b\ge 0,
\ee
and
\be
f= \pi-\arctan\fr G2+\arctan\fr{L}{hb},
\qquad  {\rm if}
 \quad b\le 0,
\ee
 with
 \be
 L =q+\fr g2b,
\ee
\ses\\
is called
the {\it  Finsleroid--Finsler  metric function}.

\ses

\ses

The  function $K$ has been normalized such that
$$
0\le f\le \pi,
$$
\ses
$$
f=0,\quad {\rm if} \quad q=0 \quad {\rm and} \quad b>0;
\qquad
f=
\pi,\quad {\rm if} \quad q=0 \quad {\rm and} \quad b<0,
$$
and
the Finsleroid length of the  vector $b^i$
is equal to 1:
\be
K(x,b^i(x))=1,
\ee
or
$$
|| b||_{\text{Finsleroid}}=1.
$$

Sometimes it is convenient to use also the function
\be
A=b+\fr g2q.
\ee
The identities
\be
L^2+h^2b^2=B, \qquad A^2+h^2q^2=B
\ee
are valid.

The zero--vector $y=0$ is excluded from consideration.
The positive (not absolute) homogeneity holds:
$$
K(x,\la y)=\la K(x,y), \qquad \la>0, ~ \forall x, ~ \forall y.
$$

{

\ses

\ses

 {\large  Definition}.  The arisen  space
\be
\cF\cF^{PD}_g :=\{\cR_{N};\,b_i(x);\,g(x);\,K(x,y)\}
\ee
is called the
 {\it Finsleroid--Finsler space}.

\ses

 {\large  Definition}. The space $\cR_N$ entering the above definition is called the {\it associated Riemannian space}.

\ses\ses

{\large Definition}.\, Within  any tangent space $T_xM$, the Finsleroid--metric function $K(x,y)$  produces the {\it Finsleroid}
 \be
 \cF^{PD}_{g\,\{x\}}:=\{y\in  \cF^{PD}_{g\,\{x\}}: y\in T_xM , K(x,y)\le 1\}.
  \ee

 \ses

{\large Definition}.\, The {\it Finsleroid Indicatrix}
 $I^{PD}_{g\,\{x\}}\in T_xM$ is the boundary of the Finsleroid:
 \be
I^{PD}_{g\,\{x\}} :=\{y\in I^{PD}_{g\,\{x\}}: y\in T_xM, K(x,y)=1\}.
  \ee

 \ses

 Since at $g=0$ the  $\cF\cF^{PD}_g$--space is
Riemannian, then the body $  \cF^{PD}_{g=0\,\{x\}}$ is a unit ball and $
I^{PD}_{g=0\,\{x\}}$ is a unit sphere.

\ses

 {\large  Definition}. The scalar $g(x)$ is called
the {\it Finsleroid charge}.
The 1-form $b=b_i(x)y^i$ is called the  {\it Finsleroid--axis}  1-{\it form}.

\ses

Under these conditions,
 we can explicitly calculate from the function $K$ the  distinguished Finslerian tensors,
 and first of all
the covariant tangent vector $\hat y=\{y_i\}$,
the  Finslerian metric tensor $\{g_{ij}\}$
together with the contravariant tensor $\{g^{ij}\}$ defined by the reciprocity conditions
$g_{ij}g^{jk}=\de^k_i$, and the  angular metric tensor
$\{h_{ij}\}$, by making  use of the following conventional  Finslerian  rules in succession:
\be
y_i :=\fr12\D{K^2}{y^i}, \qquad
g_{ij} :
=
\fr12\,
\fr{\prtl^2K^2}{\prtl y^i\prtl y^j}
=\fr{\prtl y_i}{\prtl y^j}, \qquad
h_{ij} := g_{ij}-y_iy_j\fr1{K^2}.
\ee

{

It is convenient to use the variables
\be
v^i~:=y^i-bb^i, \qquad v_m~:=u_m-bb_m=r_{mn}y^n\equiv r_{mn}v^n\equiv a_{mn}v^n,
\ee
where
 $r_{mn}=a_{mn}-b_mb_n$ is the projection tensor
(1.6). Notice that
\be
r^i{}_n~:=a^{im}r_{mn}=\de^i{}_n-b^ib_n=\D{v^i}{y^n},
\ee
\ses
\be
v_ib^i=v^ib_i=0, \qquad r_{ij}b^j=r^i{}_jb^j=b_ir^i{}_j=0
\ee
(cf. (1.7)),  and
\be
q=\sqrt{r_{ij}v^iv^j},
\ee
together with
\be
\D b{y^i}=b_i, \qquad \D q{y^i}=\fr{v_i}q, \qquad \D{(b/q)}{y^i}=\fr{2B}{NKgq^2}A_i,
\ee
where $A_i$ is the vector defined below in (A.34).

In terms of the variables (A.18) we obtain the representations
\be
y_i=\Bigl(v_i+(b+gq)b_i\Bigr)\fr{K^2}B,
\ee
\ses
\be
g_{ij}=
\biggl[a_{ij}
+\fr g{B}\Bigl (q(b+gq)b_ib_j+q(b_iv_j+b_jv_i)-b\fr{v_iv_j}q\Bigr)\biggr]\fr{K^2}B,
\ee
and the reciprocal components $(g^{ij})=(g_{ij})^{-1}$  read
\be
g^{ij}=
\biggl[a^{ij}+\fr gq(bb^ib^j-b^iy^j-b^jy^i)+\fr g{Bq}(b+gq)y^iy^j
\biggr]\fr B{K^2},
\ee
or
\be
g^{ij}=
\biggl[a^{ij}+\fr gB\Bigl(-bqb^ib^j-q(b^iv^j+b^jv^i)+(b+gq)\fr{v^iv^j}q\Bigr)
\biggr]\fr B{K^2}.
\ee
It is a useful exercise to verify that
\be
g_{ij}g^{jn}=\de^n_i.
\ee

We have also
\be
g_{ij}(x,y)\bigl|_{g=0}\bigr.= a_{ij}(x)
\ee
and
\be
g_{ij}(x,b^n(x))= a_{ij}(x),
\ee

{

together with
\be
y_ib^i=(b+gq)\fr{K^2}B, \qquad
g_{ij}b^j=\bigl(b_i+gq\fr{y_i}{K^2}\bigr)\fr{K^2}B,
\ee
\ses
\be
g_{ij}v^j=\bigl(S^2v_i+gq^3b_i\bigr)\fr{K^2}{B^2},
\ee
\ses
\be
g^{ij}a_{ij}=\fr{NB+gq^2}{K^2},
\qquad
h_{ij}b^j=\bigl(b_i-b\fr{y_i}{K^2}\bigr)\fr{K^2}B.
\ee

The determinant of the metric tensor  is everywhere positive:
\be
\det(g_{ij})=\lf(\fr{K^2}B\rg)^N\det(a_{ij})>0.
\ee

From the last expression we can explicate the vector
\be
 A_i~:=K\D{\ln\lf(\sqrt{\det(g_{mn})}\rg)}{y^i},
  \ee
obtaining
\be
A_i=\fr {NK}2g\fr1{q}(b_i-\fr b{K^2}y_i),
\ee
or
\be
A_i=\fr {NK}2g\fr1{qB}(q^2b_i- bv_i).
\ee
Raising the index according to the general rule (namely $A^i=g^{ij}A_j$)
 yields
\be
A^i=\fr N2g\fr 1{qK}
\Bigl[Bb^i-(b+gq)y^i\Bigr],
\ee
or
\be
A^i=\fr N2g\fr 1{qK}
\Bigl[q^2b^i-(b+gq)v^i\Bigr].
\ee
We have
\be
A_ib^i=\fr N2gq\fr KB, \qquad    A^ib_i=\fr N2gq\fr 1K,
\ee
so that
\be
A_hA^h=\fr{N^2}{4}g^2,
\ee

{

After that, we can elucidate the algebraic structure of the associated  Cartan tensor
\be
A_{ijk} := \fr K2\D{g_{ij}}{y^k},
\ee
which leads to
the following simple and remarkable result:
the Cartan tensor associated with the  Finsleroid--Finsler  metric function $K$
 is of the following special algebraic form:
\be
A_{ijk}=\fr1N\lf(h_{ij}A_k+h_{ik}A_j+h_{jk}A_i-\fr1{A_hA^h}A_iA_jA_k\rg).
\ee

Since
$$
\fr{v^iv^j}q\to 0 \quad {\text{when}}\quad v^i\to 0
$$
(notice (A.18)) the components  $g_{ij}$ and $g^{ij}$ given by (A.24) and (A.25)
 are smooth on all the slit tangent bundle $TM\backslash{0}$.
However, the  components of the Cartan tensor are singular at $v^i=0$, as this is apparent from the above
formulas (A.35)--(A.36) in which the {\it pole singularity} takes place at $q=0$. Therefore,  {\it
on the slit tangent bundle
$TM\backslash 0$
the $\cF\cF^{PD}_g $--space is smooth
 of the class $C^2$ and not of the class $C^3$}.

\ses


{

 We use the Riemannian metric tensor $a_{ij}$ of the associated Riemannian space
 to construct the Riemannian Christoffel symbols
\be
a^k{}_{ij}~:=\fr12a^{kn}(\prtl_ja_{ni}+\prtl_ia_{nj}-\prtl_na_{ji})
\ee
($\prtl_j=\prtl/\prtl x^j$)
which  give rise to
the Riemannian curvature tensor
\be
a_n{}^i{}_{km}=\D{a^i{}_{nm}}{x^k}-\D{a^i{}_{nk}}{x^m}+a^u{}_{nm}a^i{}_{uk}-a^u{}_{nk}a^i{}_{um}
\ee
and  the Riemannian covariant derivative
\be
\nabla_ib_j~:=\partial_ib_j-b_ka^k{}_{ij}
\ee
of the involved vector field  $b_i(x)$.
The vanishing
\be
b^j\nabla_ib_j=0
\ee
(the vector $b_j$ is of the unit Riemannian length according to  (1.1))
nullifies many terms which appear  while performing various calculations.

According to the conclusions drawn in the previous work [11-12],
the equality
\be
\nabla_ib_j=kr_{ij}
\ee
is characteristic of the property that the Finsleroid space be of the Landsberg type.
We call the  $k$ thus appeared   the {\it Finsleroid--Landsberg factor}.

{

From (A.47) we get the commutator
$$
(\nabla_m\nabla_n-\nabla_n\nabla_m)b_k=\wt k_m(a_{nk}-b_nb_k)-\wt k_n(a_{mk}-b_mb_k)
= \wt k_m r_{nk} - \wt k_nra_{mk},
$$
so that
\be
b_ja_n{}^j{}_{km}=-\wt k_kr_{nm}+\wt k_mr_{nk}
\ee
is valid with the vector
\be
\wt k_n=\D k{x^n}+k^2b_n.
\ee
In turn,
 from (A.48) it follows that
$$
(\nabla_ib_j) a_n{}^j{}_{km} + b_j \nabla_ia_n{}^j{}_{km}=
-(\nabla_i\wt k_k)   r_{nm}  +\wt k_k \nabla_i(b_nb_m)  +
(\nabla_i\wt k_m )r_{nk}  -\wt k_m \nabla_i(b_nb_k).
$$
We get
$$
 b_j \nabla_ia_n{}^j{}_{km}=
-kr_{ij} a_n{}^j{}_{km}
$$

\ses

$$
-(\nabla_i\wt k_k)   r_{nm}  +k\wt k_k (b_nr_{im}+b_mr_{in})+
(\nabla_i\wt k_m )r_{nk}  - k\wt k_m (b_nr_{ik}+b_kr_{in}).
$$
Applying here the  cyclic identity yields
$$
-kr_{ij} a_n{}^j{}_{km}
-(\nabla_i\wt k_k)   r_{nm}  +k\wt k_k (b_nr_{im}+b_mr_{in})+
(\nabla_i\wt k_m )r_{nk}  - k\wt k_m (b_nr_{ik}+b_kr_{in})
$$

\ses

$$
-kr_{kj} a_n{}^j{}_{mi}
-(\nabla_k\wt k_m)   r_{ni}  +k\wt k_m (b_nr_{ki}+b_ir_{kn})+
(\nabla_k\wt k_i )r_{nm}  - k\wt k_i (b_nr_{km}+b_mr_{kn})
$$

\ses

\be
-kr_{mj} a_n{}^j{}_{ik}
-(\nabla_m\wt k_i)   r_{nk}  +k\wt k_i (b_nr_{mk}+b_kr_{mn})+
(\nabla_m\wt k_k )r_{ni}  - k\wt k_k (b_nr_{mi}+b_ir_{mn})
=0,
\ee
\ses\\
which makes us conclude that the framework developed must be compatible with the identity
\be
\nabla_m\wt k_k -\nabla_k\wt k_m +k\bigl(\wt k_kb_m-\wt k_mb_k)=0.
\ee
It is immediately seen that special case, as given by the warped representation (1.24) of the
square $(ds)^2$, does obey the identity (A.51).

{

One considers    the {\it Finslerian Christoffel symbols}
\be
\ga^k{}_{ij}~:=g^{kn}\ga_{inj}
\ee
with
\be
\ga_{inj}~:=\fr12(\prtl_jg_{ni}+\prtl_ig_{nj}-\prtl_ng_{ji})
\ee
and construct the induced {\it spray coefficients}
\be
 G^k=\ga^k{}_{ij}y^iy^j.
\ee
The second--degree positive homogeneity
\be
G^k(x,\la y)= \la^2G^k(x,y), \qquad \la>0, ~ \forall x, ~ \forall y,
\ee
is valid.
With these coefficients,
 we obtain the  coefficients
\be
 G^i{}_k:~=\D{ G^i}{y^k}, \qquad  G^i{}_{km}:~=\D{ G^i{}_k}{y^m},
\qquad  G^i{}_{kmn}:~=\D{ G^i{}_{km}}{y^n},
\ee
and
\be
\bar G^i=\fr12G^i, \qquad \bar  G^i{}_k =\fr12 G^i{}_k, \qquad  \bar G^i{}_{km}  =\fr12G^i{}_{km},
\qquad  \bar G^i{}_{kmn}=\fr12G^i{}_{kmn}.
\ee
The homogeneity (A.55)       entails the identities
\be
2G^i= G^i{}_ky^k, \qquad  G^i{}_k=G^i{}_{km}y^m,
\qquad  G^i{}_{kmn}y^n=0.
\ee

The pair $(x,y)$, --- the so--called {\it line element}, --- is the argument of the Finslerian objects.

{

To evaluate   the  curvature tensor $R^i{}_k$, we use the well--known formula
\be
K^2R^i{}_k~:=
2\D{\bar G^i}{x^k}-\D{\bar G^i}{y^j}\D{\bar G^j}{y^k}
-y^j\Dd{\bar G^i}{x^j}{y^k}
+2\bar G^j\Dd{\bar G^i}{y^k}{y^j}
\ee
(which is tantamount to the definition (3.8.7) on p. 66 of the book [5];
$\bar G^i=
\frac12\ga^i{}_{nm}y^ny^m$,
with the Finslerian Christoffel symbols
$
\ga^i{}_{nm}$).
The concomitant tensors
\be
R^i{}_{km}~:=\fr1{3K}\Biggl(\D{(K^2R^i{}_{k})}{y^m}-\D{(K^2R^i{}_{m})}{y^k}
\Biggr)
\ee
and
\be
R_n{}^i{}_{km}~:=\D{(KR^i{}_{km})}{y^n}
\ee
arise.

\ses

The cyclic identity
\be
R_j{}^i{}_{kl|t}+R_j{}^i{}_{lt|k}+R_j{}^i{}_{tk|l}=
P_j{}^i{}_{ku}R^u{}_{lt}+P_j{}^i{}_{lu}R^u{}_{tk}+P_j{}^i{}_{tu}R^u{}_{kl}
\ee
(the formula (3.5.3) on p. 58 of the book [5]) is valid in any Finsler space.
If we contract the identity by $g^{jl}\de^k{}_i$, we get:
\be
g^{jl}\Bigl(R_j{}^i{}_{il|t}+R_j{}^i{}_{lt|i}+R_j{}^i{}_{ti|l}\Bigr)=
P^l{}^i{}_{iu}R^u{}_{lt}+P^l{}^i{}_{lu}R^u{}_{ti}+P^l{}^i{}_{tu}R^u{}_{il}.
\ee
Under the Landsberg condition, the tensor $P_{ijkl}$ is symmetric in all of its four indices and, therefore,
the  identity (A.63) reduces to merely
\be
g^{jl}\Bigl(R_j{}^i{}_{il|t}+R_j{}^i{}_{lt|i}+R_j{}^i{}_{ti|l}\Bigr)=0
\ee
($|i$ stands for the Finslerian covariant derivative),
which can be formulated as follows.

\ses\ses

{\large Proposition A1.} {\it  In any Landsberg case of
a Finsler space,
the tensor
\be
\rho_{ij}~:=\fr12(R_i{}^m{}_{mj}+R^m{}_{ijm})-\fr12g_{ij}R^{mn}{}_{nm}
\ee
is covariantly conserved:
\be
\rho^i{}_{j|i} \equiv 0.
\ee
}

\ses

It is obvious that the tensor $\rho_{ij}$, as well as  the curvature tensor
$R_n{}^i{}_{km}$,
is positively homogeneous of the degree zero with respect to the vectors $y$:
\be
\rho_{ij}(x,\la y)= \rho_{ij}(x,y), \qquad \la>0, ~ \forall x, ~ \forall y.
\ee

The explicit representation of the curvature tensor
$R_n{}^i{}_{km}$ of the Finsleroid--Landsberg--case  space  $\cF\cF^{PD}_g $
will be evaluated in detail
in Appendix B.

\ses

{

On the basis  of the  above  coefficients (A.56)--(A.58),
the Finslerian          {\it connection coefficients}
$\Ga^k{}_{ij}=\Ga^k{}_{ij}(x,y)$
are constructed according to the well--known  conventional rule:
\be
\Ga^k{}_{ij}=\ga^k{}_{ ij}-\bar G^n{}_i\3Cnkj-\bar G^n{}_j\3Cnki+\bar G^{kn}C_{nij}
\ee
with
\be
\bar G^n{}_i=\ga^n{}_{ ij}y^j-2\bar G^m\3Cmni=\Ga^n{}_{ij}y^j=\fr12 G^n{}_i
\ee
and
\be
2\bar G^m=\ga^m{}_{ij}y^iy^j=
\bar G^m{}_iy^i=\Ga^m{}_{ ij}y^iy^j=G^m,
\ee
where
 $C_{nij}=A_{nij}K^{-1}$  and $\3Cnkj=\3Ankj K^{-1}$.
By the help of these coefficients
 the $h$-{\it covariant derivatives} of tensors are constructed
as exemplified by
\be
\rho_{i|j}~:= \prtl_j \rho_i-\bar G^k{}_j\D{\rho_i}{y^k}-\Ga^k{}_{ij} \rho_k,
\ee
\ses
\be
\rho^i_{|j}~:= \prtl_j \rho^i-\bar G^k{}_j\D{\rho^i}{y^k}+\Ga^i{}_{kj} \rho^k,
\ee
\ses
\be
\rho^i{}_{k|j}~:= \prtl_j \rho^i{}_{k} -\bar G^n{}_j \D{\rho^i{}_{k}}{y^n}
+\Ga^i{}_{ nj} \rho^n{}_k-\Ga^h{}_{ kj} \rho^i{}_h,
\ee
and
\be
g_{mn|j} =
 \prtl_j  g_{mn} -\bar G^h{}_j\D{ g_{mn}}{y^h}-\Ga^h{}_{ mj}g_{hn}   -\Ga^h{}_{ nj}g_{mh},
\ee
where again $\prtl_j=\prtl/\prtl x^j$.
The coefficients (A.68) are {\it symmetric:}
\be
\Ga^k{}_{ij}=\Ga^k{}_{ji}.
\ee
The importance of the covariant derivative $|j$ thus introduced is the property that the derivative is
{\it metric},
that is we have the following identical vanishing:
\be
K_{|j}=0,  \qquad y^i_{|j}=0, \qquad l^i_{|j}=0,
\ee
and
\be
g_{mn|j}=0.
\ee
The inverse implication can readily be made, namely the conditions (A.75)--(A.77) entail uniquely the coefficients
(A.68)--(A.70).
We apply the above covariant derivative $|i$ in the Finsleroid approach developed. The derivative $|i$
extends the Riemannian covariant derivative, such that
\be
\bigl(|i\bigr)_{|g=0}=\nabla_i \quad  \text{and} \quad   \bigl(\Ga^k{}_{ij}\bigr)_{|g=0} = a^k{}_{ij}.
\ee

{

If we apply the derivative definition (A.71)  to the vector field
$b_i(x)$,
and the definition (A.72)   to the vector field  $b^i(x)$,
then, because the fields are independent of $y$, we obtain merely
\be
b_{i|j}~:= \prtl_j b_i-\Ga^m{}_{ij} b_m
\ee
and
\be
b^i_{|j}~:= \prtl_j b^i+\Ga^i{}_{mj} b^m.
\ee
Therefore,
$$
b_{|j}=y^i\Bigl(\prtl_j b_i-\Ga^m{}_{ij} b_m\Bigr)=   \prtl_j b-\bar G^m{}_jb_m
$$
(the formula (A.69) has been used), or
\be
b_{|j}=  \nabla_jb-\Bigl(\bar G^m{}_j-  a^m{}_{jk}y^k  \Bigr)b_m.
\ee

If we apply   the formulas (B.18) and (B.6)  of Appendix B,
we just conclude that the last term in the right--hand part of the last equality (A.81) vanishes, so that the following
proposition is valid.

\ses\ses

{\large Proposition A2.} {\it  In any Landsberg case of
the $\cF\cF^{PD}_g $--space,
the covariant derivatives of the 1-form $b=b_i(x)y^i$ with respect to the Finsler connection
and with respect to the associated Riemannian connection are equal to one another}:
\be
b_{|j}=\nabla_jb.
\ee

\ses\ses

Since
$$
b_{|j}=y^n\nabla_jb_n=kv_j
$$
(see (1.8)), we have
\be
b_{|j}=kv_j.
\ee

Considering the associated Riemannian metric function
\be
S=\sqrt{a_{mn}y^my^n},
\ee
we have
$
(S^2)_{|j}=-2gq \nabla_j  b
$
(use (B.20)),
that is,
\be
(S^2)_{|j}=-2kgq v_j.
\ee
Thus,
the following proposition is valid.

\ses\ses

{\large Proposition A3.} {\it  In any Landsberg case of
the $\cF\cF^{PD}_g $--space,
the covariant derivatives of the square of the associated Riemannian metric function
 with respect to the Finsler connection
is given by} (A.85).

\ses \ses

{

Taking into account
\be
S^2=q^2+b^2,
\ee
from (A.83) and (A.85) we may conclude that
\be
q_{|k}=  -\fr1q(b+gq)\nabla_kb.
\ee
Next, we get
\be
(bq)_{|k}=\fr1q
[q^2-b(b+gq)]\nabla_kb, \qquad (q/b)_{|k}=-\fr{B}{qb^2}\nabla_kb,  \qquad (b/q)_{|k}=\fr{B}{q^3}
\nabla_kb,
\ee
and
\be
J_{|n}=\fr g{2q}J\nabla_nb, \qquad B_{|n}=-\fr{gB}{q}
\nabla_nb.
\ee

This leads to
\be
K b_{|k}l^k=kq^2, \qquad   {\la}_{|k}l^k=\fr{B}{q^3} b_{|k}l^k
\ee
(where $\la=b/q$),
and
\be
S S_{|k}l^k=-gq  b_{|k}l^k, \quad  q_{|k}l^k=-\fr1q(b+gq) b_{|k}l^k,
\ee
\ses
\be
 (bq)_{|k}l^k=\fr1q \lf[q^2-b(b+gq)\rg]  b_{|k}l^k,
\quad  B_{|k}l^k=-\fr{gB}{q}  b_{|k}l^k,
\ee
where $l^k=y^k/K$ is  the unit tangent vector.

\ses

{

From (A.80) it ensues that
\be
\Bigl(b^i_{|j}\Bigr)_{|_{y^h=b^h}}       =\nabla_jb^i+\De^i{}_{mj} b^m
\ee
with the {\it osculating deflection coefficients}
\be
\De^i{}_{mj}=\De^i{}_{mj}(x)=
\Ga^i{}_{mj} (x,b(x))-a^i{}_{mj}(x),
\ee
where
$a^i{}_{mj}(x)$  are the Riemannian Christoffel symbols
given by (A.43).

Let us take the derivative (A.72) over the field $y^h=b^h(x)$.
Introducing the {\it osculating field}
\be
\rho^i_{\{b\}}=\rho^i_{\{b\}}(x)=\rho^i_{|_{y^h=b^h}}\equiv \rho^i(x,b(x)),
\ee
from (A.72), (A.93), and (A.95)  it follows that
\be
\Bigl(\rho^i_{|j}\Bigr)_{|_{y^h=b^h}}= \nabla_j \rho^i_{\{b\}}
-\Bigl(   \nabla_jb^k+\De^k{}_{mj} b^m    \Bigr)     \lf(\D{\rho^i}{y^k}\rg)_{|_{y^h=b^h}}
+\De^i{}_{kj} \rho^k_{\{b\}}.
\ee

{

In the Landsberg case of the space  $\cF\cF^{PD}_g $ under consideration, we have
\be
\nabla_jb^k=k(\de^k_j-b^kb_j)
\ee
(see (1.8))
and
\be
\De^i{}_{kj} b^k=0
\ee
(see  (B.6) and (B.20) in Appendix B).
If also  $\rho^i$ is positively homogeneous of the degree one with respect to the vectors $y$:
\be
\rho^i(x,\la y)=\la \rho^i(x,y), \qquad \la>0, ~ \forall x, ~ \forall y,
\ee
then the identity
\be
y^k\D{\rho^i}{y^k}=\rho^i
\ee
holds,
entailing
\be
b^k \lf(\D{\rho^i}{y^k}\rg)_{|_{y^h=b^h}}=\rho^i_{\{b\}}.
 \ee
Under these conditions, the representation (A.96) reduces to the equality
\be
\Bigl(\rho^i_{|j}\Bigr)_{|_{y^h=b^h}}= \nabla_j \rho^i_{\{b\}}
-k  \Biggl[   \lf(\D{\rho^i}{y^j}\rg)_{|_{y^h=b^h}} -b_j\rho^i_{\{b\}}  \Biggr]
+\De^i{}_{kj} \rho^k_{\{b\}},
\ee
which can also be written as
\be
\Bigl(\rho^i_{|j}\Bigr)_{|_{y^h=b^h}}= \nabla_j \rho^i_{\{b\}}
-k   \lf(\D{\bigl(\rho^i/b\bigr)}{y^j}\rg)_{|_{y^h=b^h}}
+\De^i{}_{kj} \rho^k_{\{b\}}.
\ee
Contracting here the indices yields
\be
\Bigl(\rho^i_{|i}\Bigr)_{|_{y^h=b^h}}= \nabla_i \rho^i_{\{b\}}
-k   \lf(\D{\bigl(\rho^i/b\bigr)}{y^i}\rg)_{|_{y^h=b^h}}
+\De^i{}_{ki} \rho^k_{\{b\}}.
\ee

{

In the particular case when the field $\rho^i_{\{b\}}$ is a factor of $b^i$, that is when
\be
\rho^i_{\{b\}}=  \ga_{\{b\}} b^i, \qquad    \ga_{\{b\}}=\ga_{\{b\}}(x),
\ee
the vanishing (A.98) suppresses the last term in (A.102), giving
\be
\Bigl(\rho^i_{|i}\Bigr)_{|_{y^h=b^h}}= \nabla_i \rho^i_{\{b\}}
-     k   \lf(\D{\bigl(\rho^i/b\bigr)}{y^i}\rg)_{|_{y^h=b^h}}.
\ee

Thus we have

\ses\ses

{\large Proposition A4.} {\it With the conditions }
 (A.97)--(A.105)
 {\it  fulfilled, the Finsler--covariant  conservation law
\be
 \rho^i_{|i}=0
\ee
 entails the Riemannian--divergence  law
\be
 \nabla_i \rho^i_{\{b\}} =    k   \lf(\D{\bigl(\rho^i/b\bigr)}{y^i}\rg)_{|_{y^h=b^h}}
\ee
  osculating along   the fundamental vector field $b^i(x)$.
}

\ses\ses

Notice that with (A.97) and (A.105) we can obtain the equality
\be
 \nabla_i \rho^i_{\{b\}}  =b^i\partial_i \ga_{\{b\}}+(N-1) k  \ga_{\{b\}}.
\ee

\ses

Similar procedure of projecting on the vector field $b^i(x)$
can be addressed to the conservation law of the form $\rho^i{}_{j|i}=0$.

{

In terms of the $b$--geodesic coordinates $\{z^A\}$, evaluation of the components of the
Christoffel symbols (A.43) on the basis of the warped metric function (1.14)
yields the  following list:
\be
s^N{}_{NN}=0,\qquad s^a{}_{NN}=0,\qquad s^N{}_{bN}=0,
\ee
\ses
\be
s^a{}_{bN}=\fr12r^{ac}\D{r_{bc}}{z^N},\qquad s^N{}_{bc}=-\fr12\D{r_{bc}}{z^N},
\qquad s^a{}_{bc}=\fr12r^{ad}\Biggl(\D{r_{dc}}{z^b}+\D{r_{bd}}{z^c}-\D{r_{bc}}{z^d}\Biggr).
\ee
The equalities $ b_N=1$ and $ b_a=0$ entail
\be
(\nabla b)_{NN}=0,
\ee
\ses
\be
(\nabla b)_{Na}=  0,
\ee
\ses
\be
 (\nabla b)_{aN}= 0,
\ee
and
 \be
(\nabla b)_{ab}=\fr12\D{r_{ab}}{z^N}.
\ee
When we apply the factorization
\be
r_{ab}(z^A)=\lf(\phi(z^A)\rg)^2p_{ab}(z^c),
\ee
so that
\be
(ds)^2=\lf(dz^N\rg)^2+\lf(\phi(z^A)\rg)^2p_{ab}(z^c)dz^adz^b,
\ee
we obtain
\be
\D {\phi}{z^N}=k\phi
\ee
and (A.115) reduces to
 \be
(\nabla b)_{ab}=kr_{ab}.
\ee

{

The positive--definite $\cF\cF^{PD}_g $--space described
possesses
the indefinite (relativistic) version, to be denoted as  the $\cF\cF^{SR}_g $--space
(with the upperscripts  ``SR'' meaning ``special--relativistic'').
The transition from the first space to the second space  implies the formal change
$$
g\to ig
$$
of the Finsleroid parameter $g$,
where $i$ stands for the imaginary unity.
The underlined space $\cR_N=\{M, a_{mn}\}$ is now taken to be {\it pseudo-Riemannian},
such that   the input metric tensor $\{a_{mn}(x)\}$  is to be
 pseudo--Riemannian with
the {\it time--space signature:}
\be
{\rm sign}(a_{mn})=(+ - -\dots).
\ee

{

Generalizing  accordingly the pseudo--Riemannian geometry in a pseudo--Finsleroid Finslerian way,
we are to adapt the consideration to the  following decomposition of the tangent bundle $TM$:
\be
TM=\cS_g^+\cup \Si_g^+\cup{\cR_g}\cup\Si_g^-\cup\cS_g^-,
\ee
which sectors relate to the cases that the tangent vectors $y\in TM$ are, respectively,
time--like, upper--cone isotropic, space--like, lower--cone isotropic, or past--like.
The sectors  are defined according to the following list:
\be
\cS_g^+=
\Bigl(y\in \cS_g^+:~y\in T_xM,\,  b(x,y)>-g_-(x)q(x,y)\Bigr),
\ee
\medskip
\be
\Si_g^+=\Bigl(y\in \Si_g^+:~y\in T_xM, \, b(x,y)=-g_-(x)q(x,y)\Bigr),
\ee
\medskip
\be
\cR_g^+=
\Bigl(y\in \cR_g^+:~y\in T_xM,\,  -g_-(x)q(x,y)>b(x,y) >0 \Bigr),
\ee
\medskip
\be
\cR^0=
\Bigl(y\in \cR^0:~y\in T_xM,\,  b(x,y) =0 \Bigr),
\ee
\medskip
\be
\cR_g^-=
\Bigl(y\in \cR_g^-:~y\in T_xM,\,0>b(x,y) >-g_+(x)q(x,y) \Bigr),
\ee
\medskip
\be
\Si_g^-=\Bigl(y\in \Si_g^-:~y\in T_xM, \, b(x,y)=-g_+(x)q(x,y)\Bigr),
\ee
\medskip
\be
\cS_g^-=
\Bigl(y\in \cS_g^-:~y\in T_xM,\,  b(x,y)<-g_+(x)q(x,y)\Bigr),
\ee
\medskip
\be
\cR_g=\cR_g^+\cup\cR_g^-\cup\cR^0.
\ee

{

We use the convenient notation
\be
G= \fr gh,
\qquad h = \sqrt{1+\fr14g^2}
\ee
(instead of (A.2)),
\be
g_+=-\fr12g+h, \qquad g_-=-\fr12g-h,
\ee
\medskip
\be
G_+=\fr{g_+}h\equiv -\fr12G+1, \qquad G_-=\fr{g_-}h\equiv -\fr12G-1,
\ee
\medskip
\be
g^+=\fr 1{g_+}=-g_-,  \qquad  g^-=\fr 1{g_-}=-g_+,
\ee
\medskip
\be
g^+=\fr12g+h, \qquad g^-=\fr12g-h,
\ee
\medskip
\be
G^+=\fr{g^+}h\equiv \fr12G+1, \qquad G^-=\fr{g^-}h\equiv \fr12G-1.
\ee
The following identities hold
\be
g_++g_-=-g, \qquad g_+-g_-=2h,
\ee
\medskip
\be
g^++g^-=g, \qquad g^+-g^-=2h,
\ee
\medskip
\be
g_+g_-=-1, \qquad
g^+g^-=-1,
\ee
together with
the {\it
$g$--symmetry}
\be
g_+\g -g_-, \qquad g^+\g -g^-, \qquad G_+\g -G_-, \qquad G^+\g -G^-.
\ee

{

It is implied that $g=g(x)$ is a scalar on the underlying manifold $M$.
All the range
\be
-\infty <g(x)<\infty
\ee
(instead of (A.1)) is now
admissible.
We also  assume that the manifold $M$ admits
a   1-form $b=b(x,y)$ which is
{\it timelike} in terms of the pseudo--Riemannian metric ${\cal S}$,
 such that the pseudo--Riemannian length of the involved vector $b_i$   be equal to 1.
With respect to  natural local coordinates in the space
$\cR_N$
we have the local representations
\be
 b=b_i(x)y^i,
\ee
\ses
\be
S= \sqrt{|a_{ij}(x)y^iy^j|},
\ee
\ses
\be
q=\sqrt{|r_{ij}(x)y^iy^j|},
\ee
\ses
\be
r_{ij}(x)=b_i(x)b_j(x)-a_{ij}(x),
\ee
\ses
\be
a^{ij}b_ib_j=1,
\ee
\ses
\be S^2=b^2-q^2,
\ee
\ses
\be
b^ir_{ij}=0,
\ee
where
\be
b^i~:=a^{ik}b_k
\ee
(compare with (A.18)--(A.21)).

{

The {\it  pseudo--Finsleroid characteristic
quadratic form}
\be
B(x,y) :=b^2-gqb-q^2
\equiv (b+g_+q)(b+g_-q)
\ee
is now of the positive discriminant
\be
D_{\{B\}}=4h^2>0
\ee
(compare these formulas with (A.3) and (A.4)).

In terms of these concepts, we propose

\ses

{\large Definition}. The scalar function $F(x,y)$ given by the formula
\be
F(x,y)~:=\sqrt{|B(x,y)|}\,J(x,y)
\equiv
|b+g_-q|^{G_+/2}|b+g_+q|^{-G_-/2},
\ee
where
\be
J(x,y)=
\lf|
\fr{b+g_-q}{b+g_+q}
\rg|^{-G/4},
\ee
is called
the {\it  pseudo-Finsleroid--Finsler  metric function}.

\ses
\ses

Again,
the zero--vector $y=0$ is excluded from consideration:
\be
y\ne 0.
\ee

The normalization of the above function $F$
is compatible with fulfilling  the unit property
\be
F(x,b^i(x))=1
\ee
(similarly to (A.11)).

The positive  (not absolute) homogeneity  holds:
\be
F(x,\la y)=\la F(x,y), \qquad \la>0, ~ \forall x, ~ \forall y.
\ee
The functions
 \be
 L(x,y) =q-\fr g2b
\ee
and
\be
A(x,y)=b-\fr g2q
\ee
are now to be used instead of (A.10) and (A.12), so that (A.13) changes to read
\be
L^2-h^2b^2=B, \qquad A^2-h^2q^2=B.
\ee

{

Similarly to (A.14), we introduce

\ses

 {\large Definition}.  The arisen  space
\be
\cF\cF^{SR}_g :=\{\cR_{N};\,b(x,y);\,g(x);\,F(x,y)\}
\ee
is called the
 {\it pseudo--Finsleroid--Finsler space}.

\ses

The upperscript ``SR" emphasizes the Specially  Relativistic character of the space under study.

\ses

 {\large Definition}. The space $\cR_N=(M,\,{\cal S})$ entering the above definition is called the {\it associated
pseudo--Riemannian space}.

\ses

 {\large Definition}. The scalar $g(x)$ is called
the {\it  pseudo--Finsleroid charge}.
The 1-form $b$ is called the  {\it  pseudo--Finsleroid--axis}  1-{\it form}.

\ses

It can be verified that the Finslerian metric tensor constructed from the function $F$ given by (A.151)
does inherit from the tensor $\{a_{ij}(x)\}$
the time--space signature (A.120):
\be
{\rm sign}(g_{ij})=(+ - -\dots).
\ee

The structure (A.42) for the
 Cartan tensor remains valid in the pseudo--Finsleroid case, now with
 \be
A_hA^h=-\fr{N^2}{4}g^2
\ee
(compare with (A.40)).
Elucidating the structure of
the respective indicatrix curvature tensor
 (2.29) of the  $\cF\cF^{SR}_g$--space again results
in the conclusion that the indicatrix curvature value
$ \cR_{\text{pseudo-Finsleroid  Indicatrix} }$
is
\be
 \cR_{\text{pseudo-Finsleroid  Indicatrix} }=-  \lf(1+\fr14g^2\rg) \le -1,
\ee
 so that
$$
\cR_{\text{pseudo-Finsleroid Indicatrix} }\stackrel{g\to 0}{\Longrightarrow}
\cR_{\text{pseudo-Euclidean Sphere}}=-1.
$$
{\it The pseudo--Finsleroid indicatrix
 is a space of constant  negative curvature}.

{

Again, it is convenient to use the variables
\be
v^i~:=y^i-bb^i, \qquad v_m~:=u_m-bb_m=r_{mn}y^n\equiv r_{mn}v^n\equiv a_{mn}v^n,
\ee
where
\be
r_{mn}=a_{mn}-b_mb_n
\ee
 is the projection tensor. We obtain the identities
\be
r^i{}_n~:=a^{im}r_{mn}=\de^i{}_n-b^ib_n=\D{v^i}{y^n},
\ee
and
\be
v_ib^i=v^ib_i=0, \qquad r_{ij}b^j=r^i{}_jb^j=b_ir^i{}_j=0.
\ee
It is useful to apply the notation
\be
q=\sqrt{r_{mn}v^mv^n}.
\ee
From the pseudo-Finsleroid-Finsler metric function (A.151) we straightforwardly evaluate
the explicit representations
\be
y_i=\Bigl(v_i+(b-gq)b_i\Bigr)\fr{F^2}B
\ee
and
\be
g_{ij}=
\biggl[a_{ij}
+\fr g{B}\Bigl (-q(b+gq)b_ib_j-q(b_iv_j+b_jv_i)-b\fr{v_iv_j}q\Bigr)\biggr]\fr{F^2}B.
\ee
Through  the reciprocity $(g^{ij})=(g_{ij})^{-1}$, we arrive at the representations
\be
g^{ij}=
\biggl[a^{ij}+\fr gq(bb^ib^j-b^iy^j-b^jy^i)+\fr g{Bq}(b-gq)y^iy^j
\biggr]\fr B{F^2}
\ee
and
\be
g^{ij}=
\biggl[a^{ij}+\fr gB\Bigl(bqb^ib^j+q(b^iv^j+b^jv^i)+(b-gq)\fr{v^iv^j}q\Bigr)
\biggr]\fr B{F^2}.
\ee

It is obvious that
\be
g_{ij}(x,y)\bigl|_{g=0}\bigr.= a_{ij}(x)
\ee
and
\be
g_{ij}(x,b^n(x))= a_{ij}(x).
\ee

{

We observe the phenomenon that  the representations (A.168)--(A.171)
are directly obtainable from the positive--definite case representations  (A.23)--(A.26)
through the formal change:
\be
g ~ \stackrel{PD ~ \to ~SR}{\Longrightarrow} ~ ig
\ee
and
\be
q ~ \stackrel{PD ~ \to ~SR}{\Longrightarrow} ~ iq,
\ee
where $i$ stands for the imaginary unity.
Therefore, we may apply the rules
\be
\fr gq ~ \stackrel{PD ~ \to ~SR}{\Longrightarrow} ~ \fr gq,   \qquad
 gq ~ \stackrel{PD ~ \to ~SR}{\Longrightarrow} ~  - gq.
\ee
It is the useful exercise to verify that if we apply these rules
 to the expression (A.151) of the relativistic function $F$, we obtain the positive--definite
case function $K$ defined by (A.6).

\vskip 2cm

\ses

\ses

\setcounter{equation}{0}

\nin
{\bf\large Appendix B:~  Evaluation and application of curvature tensor
}

\ses\ses

In addition to the variables
\be
v^m=  y^m-bb^m, \qquad v_n=a_{mn}v^m
\ee
(see (A.18)),
we shall use the tensor
$$
v^m{}_n~:=\D{v^m}{y^n}+\fr1{ q^2}v^mv_n,
$$
so that,
\be
v^m{}_n=r^m_n  +    \fr1{ q^2}v^mv_n
\ee
\ses
\be
b^nv^m{}_{n} =0,
\ee
and
\be
 y^n v^m{}_n=2v^m,
\ee
together with the tensor
\be
v^m{}_{hn}~:=v^m{}_hv_n+v^m{}_nv_h+v^mv_{hn}-\fr 4{ q^2}v^mv_hv_n
\ee
which fulfills the identities
\be
b^nv^m{}_{hn} =b_mv^m{}_{hn}   =0,  \qquad  y^nv^m{}_{hn} =q^2v^m{}_h.
\ee

We shall also use the notation
\be
 s_i=y^j\nabla_ib_j, \qquad s^i=a^{in}s_n,    \qquad  (ys)=y^is_i.
 \ee
These objects  don't depend on the vectors $y$.

{

Below, we shall evaluate the curvature tensor of the  $\cF\cF^{PD}_g $--space
in the Landsberg case, that is, when the conditions
\be
g=const
\ee
and
\be
\nabla_ib_n=  kr_{in}
\ee
(see (1.8) and (1.23))
are satisfied.
The formulae indicated in (B.7) reduce to
\be
s_i=kv_i, \qquad s^i=kv^i, \qquad (ys)=kq^2.
\ee

We shall denote
\be
k_i=\D k{x^i}
\ee
and
\be
(yk)=y^ik_i,
\ee
and  also
\be
\wt k_i=k_i+k^2b_i
\ee
together with
\be
(y\wt k)=y^i\wt k_i =(yk)+k^2b
\ee
and
\be
n_1=(y\wt k)-b(b\wt k).
\ee

{

The spray coefficients $G^m$ entailed have been found in the previous work [11-12]. They read merely
\be
G^m=
k g  qv^m
+a^m{}_{jn}y^jy^n.
\ee
It is notable  that  the Finsleroid--Finsler metric function $K$ does not enter the right--hand side of these
$G^m$.
The presence of the constant $g$ in the right--hand side of (B.16)
 is the only trace of the function $K$ in
these coefficients.

{

The coefficients
\be
G^m{}_n=\D{G^m}{y^n}
\ee
equal
\be
G^m{}_n=
kg  qv^m{}_n
+2a^m{}_{jn}y^j,
\ee

{

and the coefficients
\be
G^m{}_{nh}=\D{G^m_n}{y^h}
\ee
equal
\be
G^m{}_{nh}=  \fr {k g }{ q}   v^m{}_{hn}    +2a^m{}_{hn},
\ee
where   the tensor (B.5) has been used.

\ses

{

Differentiating (B.16) yields
\be
\D{G^m}{x^i}=
 g qk_i
v^m
-k g  q(s_ib^m+b\nabla_i b^m)
-
k g \fr1ws_iv^m
+\De.
\ee
Here and below, $w=q/b$ and
 $\De$ symbolizes the summary of the terms which involve partial derivatives
of the input Riemannian metric tensor $a_{ij}$ with respect to the coordinate variables $x^k$.

Simple direct calculations yield
$$
\Dd{G^m}{y^k}{x^i}= g \fr1qk_i  v_kv^m
+  g qk_i  r^m{}_k
-k g \fr1 qv_k(s_ib^m+b\nabla_i b^m)
-k g  q(b^m\nabla_kb_i+b_k\nabla_i b^m)
$$
\ses
\be
-k g \fr1{q}b_kv^ms_i
+
k g \fr b{q^3}v_kv^ms_i
-k g \fr1wv^m\nabla_kb_i-
k g \fr1ws_ir^m{}_k
+\De
\ee
\ses
and
$$
y^i\Dd{G^m}{y^k}{x^i}= g \fr1q(yk)  v_kv^m
+  g q(yk)  r^m{}_k
-k g \fr1 qv_k\Bigl((ys)b^m+bs^m\Bigr)
-k g  q(b^m s_k+b_ks^m)
$$
\ses
\be
-k g \fr1{q}b_kv^m(ys)
+
k g \fr b{q^3}v_kv^m(ys)
-k g \fr1wv^ms_k-
k g \fr1w(ys)r^m{}_k
+\De,
\ee

{

together with
$$
2\D{\bar G^i}{x^k}       -     y^j\Dd{\bar G^i}{x^j}{y^k}
=
 g qk_k
v^i
-k g  q(s_kb^i+b\nabla_k b^i)
-
k g \fr1ws_kv^i
$$

         \ses

$$
-\fr12 g \fr1q(yk)  v_kv^i
-\fr12  g q(yk)  r^i{}_k
+\fr12k g \fr1 qv_k\Bigl((ys)b^i+bs^i\Bigr)
+\fr12  k g  q(b^i s_k+b_ks^i)
$$
\ses
\be
+\fr12  k g \fr1{q}b_kv^i(ys)
-\fr12
k g \fr b{q^3}v_kv^i(ys)
+\fr12  k g \fr1wv^is_k
+\fr12
k g \fr1w(ys)r^i{}_k
+\De.
\ee

{

This entails
$$
2\D{\bar G^i}{x^k}       -     y^j\Dd{\bar G^i}{x^j}{y^k}
=
-\fr12  g q(yk) \lf( r^i{}_k+\fr1{q^2}v^iv_k\rg)
+ g qk_k
v^i
$$

\ses

$$
-\fr12 k g  qs_kb^i        -       k g  qb\nabla_k b^i
- \fr12
k g \fr1ws_kv^i
+\fr12k g \fr1 qv_k\Bigl((ys)b^i+bs^i\Bigr)
+\fr12  k g  qb_ks^i
$$

\ses

\be
+\fr12  k g \fr1{q}b_kv^i(ys)
+\fr12
k g \fr1w(ys)   \lf( r^i{}_k-\fr1{q^2}v^iv_k\rg)
+\De
\ee

{

and
\be
 G^i{}_j G^j{}_k=k^2g^2q^2   \lf(v^i{}_k+\fr 2 {q^2}v^iv_k\rg)
 +\De,
\ee
together with
\be
G^j\Dd{ G^i}{y^k}{y^j}=k^2g^2q^2 v^i{}_k
 +\De
 \ee
\ses
and
\be
-\D{\bar G^i}{y^j}\D{\bar G^j}{y^k}
+2\bar G^j\Dd{\bar G^i}{y^k}{y^j}=
\fr14 k^2g^2q^2   \lf(v^i{}_k -   \fr 2  {q^2}v^iv_k\rg)
+\De.
\ee

{

\ses

With these formulae,
we can obtain the explicit form of the curvature tensor $R^i{}_k$ on the basis of the definition (A.24) written in
the previous Appendix A.
The result reads
$$
K^2R^i{}_k
=
-\fr12  g q(yk) \lf( r^i{}_k+\fr1{q^2}v^iv_k\rg)
+ g qk_k
v^i
$$

\ses

$$
-\fr12 k g  qs_kb^i        -       k g  qb\nabla_k b^i
- \fr12
k g \fr1ws_kv^i
+\fr12k g \fr1 qv_k\Bigl((ys)b^i+bs^i\Bigr)
+\fr12  k g  qb_ks^i
$$

\ses

$$
+\fr12  k g \fr1{q}b_kv^i(ys)
+\fr12
k g \fr1w(ys)   \lf( r^i{}_k-\fr1{q^2}v^iv_k\rg)
$$

\ses

\be
+
\fr14 k^2g^2q^2   \lf(r^i{}_k -   \fr 1  {q^2}v^iv_k\rg)
+\De.
\ee

\ses

{

Using the claimed equalities
$
\nabla_mb_n=kr_{mn}, ~s_i= kv_i, ~ s^i=kv^i, ~ (ys)=kq^2
$
(see (B.9) and (B.10))
evokes much simplification, leaving us with
$$
K^2R^i{}_k
=
-\fr12  g q(yk) \lf( r^i{}_k+\fr1{q^2}v^iv_k\rg)
+ g qk_k
v^i
$$

  \ses

\be
+  k^2 g  qb_kv^i
-\fr12
k^2 gqb \lf( r^i{}_k+\fr1{q^2}v^iv_k\rg)
+
\fr14 k^2g^2q^2   \lf(r^i{}_k -   \fr 1  {q^2}v^iv_k\rg)
+\De.
\ee

{

Eventually,
\be
K^2R^i{}_k
=
\fr14 k^2g^2q^2   \lf(r^i{}_k -   \fr 1  {q^2}v^iv_k\rg)
-\fr12  g q(y\wt k) \lf( r^i{}_k+\fr1{q^2}v^iv_k\rg)
+ g q\wt k_k
v^i
+y^na_n{}^i{}_{km}y^m.
\ee

Thus we have arrived at the following proposition.

\ses\ses

{\large Proposition B1.} {\it   In the Finsleroid space
$\cF\cF^{PD}_g $, the Landsberg--case curvature tensor $R^i{}_k$
is explicitly given by the previous formula.
}

\ses

{

The entailed tensor (A.25) is found to read
$$
3KR^i{}_{km}=\fr 34 k^2g^2 (v_mr^i{}_k -v_kr^i{}_m)
$$
\ses
$$
-\fr12  g q
\Biggl[ \wt k_m  \lf( r^i{}_k+\fr1{q^2}v^iv_k\rg)-   \wt k_k \lf( r^i{}_m+\fr1{q^2}v^iv_m\rg)\Biggr]
$$
\ses

$$
+\fr g q(v_m\wt k_k-v_k\wt k_m)     v^i
+ g q(\wt k_kr^i{}_m  - \wt k_mr^i{}_k)
+3y^na_n{}^i{}_{km}
$$

\ses\ses

$$
=\fr 34 k^2g^2  (v_mr^i{}_k -v_kr^i{}_m)
$$
\ses
$$
-\fr 32  g q
\Biggl[ \wt k_m  \lf( r^i{}_k+\fr1{q^2}v^iv_k\rg)-   \wt k_k \lf( r^i{}_m+\fr1{q^2}v^iv_m\rg)\Biggr]
+3y^na_n{}^i{}_{km}.
$$
\ses

{

Simplifying yields
$$
KR^i{}_{km}
=\fr 14 k^2g^2  (v_mr^i{}_k -v_kr^i{}_m)
$$

\ses

\be
-\fr 12  g q
\Biggl[ \wt k_m  \lf( r^i{}_k+\fr1{q^2}v^iv_k\rg)-   \wt k_k \lf( r^i{}_m+\fr1{q^2}v^iv_m\rg)\Biggr]
+  y^na_n{}^i{}_{km}.
\ee

\ses

{

The full curvature tensor (A.26) is obtained to read
$$
R_n{}^i{}_{km}
=\fr 14 k^2g^2  (r_{mn}r^i{}_k -r_{kn}r^i{}_m)
$$

\ses

\ses
$$
- \fr g {2q}  v_n
\Biggl[ \wt k_m  \lf( r^i{}_k-\fr1{q^2}v^iv_k\rg)-   \wt k_k \lf( r^i{}_m-\fr1{q^2}v^iv_m\rg)\Biggr]
$$
\ses

\ses

\ses
\be
+ \fr g {2q}
\Biggl[   r^i{}_n( \wt k_k v_m - \wt k_m v_k)   +v^i(\wt k_k r_{mn}-    \wt k_m r_{kn}   )\Biggr]
+  a_n{}^i{}_{km}.
\ee

{

The tensor can also be represented in the short form
\be
R_n{}^i{}_{km}
=\fr 14 k^2g^2  (r_{mn}r^i{}_k -r_{kn}r^i{}_m)
+ \fr {g}{2q}  (\wt k_k r^i{}_{nm}-\wt k_mr^i{}_{nk})
+  a_n{}^i{}_{km},
\ee
with the help of the tensor
\be
r^i{}_{nm}= r_{nm}v^i + r^i_mv_n+r^i_nv_m -  \fr 1{q^2}v^iv_nv_m.
 \ee

\ses\ses


{

From (B.33) we obtain the contraction

\ses

$$
R_n{}^i{}_{im}  =  \fr 14 (N-2)  k^2g^2  r_{mn}
- \fr g {2q}  v_n
\Biggl[ (N-2)\wt k_m- \wt k_ir^i{}_m + \fr1{q^2}  \wt k_iv^iv_m\Biggr]
$$

\ses

\ses

$$
+ \fr g {2q}
\lf[   r^i{}_n \wt k_i v_m -\wt k_m v_n
 +v^i\wt k_i r_{mn}-    \wt k_m v_n\rg]
+  a_n{}^i{}_{im},
$$
\ses
or, after due simplification,
$$
R_n{}^i{}_{im}  =  \fr 14 (N-2)  k^2g^2  r_{mn}
- \fr {Ng} {2q}  v_n  \wt k_m
$$

\ses

\ses

\be
+ \fr g {2q} ( v_n \wt k_ir^i{}_m +   v_m \wt k_ir^i{}_n)
+ \fr g {2q}
  n_1        
 \lf( r_{mn}-  \fr1{q^2}v_mv_n\rg)
+  a_n{}^i{}_{im}.
\ee

{

Next, we find
\be
a^{nm}R_n{}^i{}_{im}  =  \fr 14 (N-2) (N-1) k^2g^2
- \fr {Ng} {2q}
  n_1              
+ \fr g {q}
  n_1        
+ \fr g {2q}(N-2)
  n_1            
+  a^{nm}a_n{}^i{}_{im}.
\ee

{

Let us now evaluate the scalar
\be
R^{ni}{}_{in}=g^{nm}R_n{}^i{}_{im},
\ee
applying the representation

$$
g^{nm}=
\biggl[a^{nm}+\fr gq(bb^nb^m-b^ny^m-b^my^n)+\fr g{Bq}(b+gq)y^ny^m
\biggr]\fr B{K^2}
$$
(see (A.25)).
We obtain
$$
\fr {K^2} B R^{ni}{}_{in}= \fr 14 (N-2) (N-1) k^2g^2    +  a^{nm}a_n{}^i{}_{im}
 - \fr {Ng} {2q}  n_1   + \fr g {q}  n_1  + \fr g {2q}(N-2)  n_1
 $$

\ses\ses

$$
+\fr{gb}qb^mb^n  R_n{}^i{}_{im}
-\fr{g}qy^mb^n  R_n{}^i{}_{im}
-\fr{g}qb^my^n  R_n{}^i{}_{im}
+  \fr g{Bq}(b+gq)    y^my^n  R_n{}^i{}_{im}
$$

{

$$
= \fr 14 (N-2) (N-1) k^2g^2
+\fr{g^2}2  N
b^m  \wt k_m
$$

\ses
\ses

$$
 - \fr {Ng} {2q}  n_1   + \fr g {q}  n_1  + \fr g {2q}(N-2)  n_1
 $$

\ses\ses

$$
+  \fr g{Bq}(b+gq)        \Biggl[
\fr 14(N-2)  k^2g^2 q^2
-\fr 12  g q
\Biggl( N (y\wt k) -   2\wt k_i v^i\Biggr)
\Biggr]
+\fr {K^2} B g^{nm}a_n{}^i{}_{im}
$$

\ses

$$
= \fr 14 (N-2) (N-1) k^2g^2
+\fr{g^2}2  N
b^m  \wt k_m
$$

\ses\ses

$$
 - \fr {Ng} {2q}  n_1   + \fr g {q}  n_1  + \fr g {2q}(N-2)  n_1
 $$

\ses\ses

$$
+  \fr g{bq} \Biggl[
\fr 14(N-2)  k^2g^2 q^2
-\fr 12  g q
\Biggl( N (y\wt k) -   2\wt k_i v^i\Biggr)
\Biggr]
$$
\ses

$$
-  \fr {gq}{Bb}   \Biggl[
\fr 14(N-2)  k^2g^2 q^2
-\fr 12  g q
\Biggl( N (y\wt k) -   2\wt k_i v^i\Biggr)
\Biggr]
+\fr {K^2} B g^{nm}a_n{}^i{}_{im}.
$$

{

Eventually, we obtain  the representation
$$
\fr {K^2} B R^{ni}{}_{in}
= (N-2)  \Biggl[
\fr 14  (N-1) k^2g^2       +  \fr 14 \fr1b k^2g^3 q
-\fr 12 \fr1b  g^2
  n_1           
\Biggr]
$$

\ses\ses

$$
 - \fr {Ng} {2q}  n_1   + \fr g {q}  n_1  + \fr g {2q}(N-2)  n_1
 $$

\ses\ses

\be
-  \fr {gq}{Bb}   \Biggl[
\fr 14(N-2)  k^2g^2 q^2
-\fr 12  g q
\Bigl( N (y\wt k)
 -  2  n_1               
 \Bigr)
\Biggr]
+\mu,
\ee
\ses
where
\be
\mu~:=\fr {K^2} B g^{nm}a_n{}^i{}_{im}.
\ee
\ses
We have
\be
\mu= \fr {K^2} B g^{nm}a_n{}^i{}_{im}= a^{nm}a_n{}^i{}_{im}
+\fr{gb}qb^mb^n  a_n{}^i{}_{im}
-2\fr{g}qy^mb^n  a_n{}^i{}_{im}
+  \fr g{Bq}(b+gq)    y^my^n  a_n{}^i{}_{im}.
\ee

{

Use the components
$$
g_{ij}=
\biggl[a_{ij}
+\fr g{B}\Bigl (q(b+gq)b_ib_j+q(b_iv_j+b_jv_i)-b\fr{v_iv_j}q\Bigr)\biggr]\fr{K^2}B
$$
(see (A.24)) to lower the index according to
\be
R_{nikm}=g_{ji}R_n{}^j{}_{km}.
\ee
\ses
We obtain
$$
\fr B{K^2}R_{nikm}
=\fr 14 k^2g^2  (r_{mn}r_{ik} -r_{kn}r_{im})
$$

\ses

\ses

$$
- \fr g {2q}  v_n
\Biggl[ \wt k_m  \lf( r_{ik}-\fr1{q^2}v_iv_k\rg)-   \wt k_k \lf( r_{im}-\fr1{q^2}v_iv_m\rg)\Biggr]
$$

\ses\ses

$$
+ \fr g {2q}
\Biggl[   r_{in}( \wt k_k v_m - \wt k_m v_k)   +v_i(\wt k_k r_{mn}-    \wt k_m r_{kn}   )\Biggr]
+\fr B{K^2}  g_{ji}a_n{}^j{}_{km}
$$

\ses\ses

\be
+\fr g{B}\Bigl (qb_i-b\fr{v_i}q\Bigr)
\Biggl[
\fr 14 k^2g^2  (r_{mn}v_k -r_{kn}v_m)
+ \fr g {2q}
\Biggl(  v_n( \wt k_k v_m - \wt k_m v_k)   +q^2(\wt k_k r_{mn}-    \wt k_m r_{kn}   )\Biggr)
\Biggr],
\ee
where
\be
\fr B{K^2}  g_{ji}a_n{}^j{}_{km}=
a_{nikm}  +  \fr g{B}   \Bigl(q(b+gq)b_i+qv_i\Bigr) b_ja_n{}^j{}_{km}
+\fr g{B}\Bigl (qb_i-b\fr{v_i}q\Bigr) v_j a_n{}^j{}_{km}.
\ee

{

Also, we must perform the involved calculation to find the contracted tensor
\be
R^m{}_{ikm} =g^{nm}R_{nikm}.
\ee
We obtain
\ses\\
$$
R^m{}_{ikm}  =\fr 14 k^2g^2 (N-2)r_{ik}
- \fr g {2q}
  v^m\wt k_m
 \lf( r_{ik}-\fr1{q^2}v_iv_k\rg)
$$

\ses

\ses

$$
+ \fr g {2q}
\Biggl[  ( \wt k_k v_i -  r_i{}^m \wt k_m v_k)   +v_i\Bigl((N-1)\wt k_k -  \wt k_m r_k{}^m \Bigr)  \Biggr]
$$

\ses\ses

$$
+\fr g{B}\Bigl (qb_i-b\fr{v_i}q\Bigr)
\Biggl[
\fr 14 k^2g^2 (N-2)v_k
+ \fr g {2q}
\Biggl(  (q^2 \wt k_k  - v^m\wt k_m v_k)   +q^2\Bigl((N-1)\wt k_k-    \wt k_m r_k^m\Bigr)   \Biggr)
\Biggr]
$$

\ses

\ses

$$
+ \fr {g^2} {2}
 b^m\wt k_m  \lf( r_{ik}-\fr1{q^2}v_iv_k\rg)
$$

         \ses            \ses

$$
+\fr g{Bq}(b+gq)
\Biggl[
\fr 14 k^2g^2q^2  \lf( r_{ik}-\fr1{q^2}v_iv_k\rg)
- \fr12 gq
(y \wt k)  \lf( r_{ik}-\fr1{q^2}v_iv_k\rg)
$$

\ses

\ses

\be
+ \fr g {q}   v_i\Bigl(q^2 \wt k_k  - (y\wt k) v_k\Bigr)
+\fr g{B}\Bigl (qb_i-b\fr{v_i}q\Bigr)
gq
\Bigl(q^2 \wt k_k  -(y\wt k) v_k\Bigr)
\Biggr]
+\theta_i,
\ee
\ses
where
\be
 \theta_i~:=  g_{ji}a_n{}^j{}_{km}g^{nm}.
\ee
\ses

{

Adding   (B.37) and (B.46) yields
\ses
$$
R^m{}_{ikm}         +   R_i{}^h{}_{hk}
=\fr 12 k^2g^2 (N-2)r_{ik}
+ \fr {g^2} {2}
 b^m\wt k_m  \lf( r_{ik}-\fr1{q^2}v_iv_k\rg)
$$

\ses    \ses

$$
+\fr g{B}\Bigl (qb_i-b\fr{v_i}q\Bigr)
\Biggl[
\fr 14 k^2g^2 (N-2)v_k
+ \fr {gq} {2}
\Biggl( \wt k_m r_k^m   -  \fr1{q^2} \wt k_mv^m v_k   - N \wt k_k \Biggr)
\Biggr]
$$

\ses

\ses

$$
+\fr g{Bq}(b+gq)
\Biggl[
\fr 14 k^2g^2q^2  \lf( r_{ik}-\fr1{q^2}v_iv_k\rg)
- \fr12 gq
(y \wt k)  \lf( r_{ik}-\fr1{q^2}v_iv_k\rg)
$$

\ses

\ses

$$
+ \fr g {q}
 v_i\Bigl(q^2 \wt k_k  - (y\wt k) v_k\Bigr)
$$

\ses

\ses

\be
+\fr {g^2q^2}{B}b_i \Bigl(q^2 \wt k_k  -(y\wt k) v_k\Bigr)
-\fr {g^2q}{B}b\fr{v_i}q \Bigl(q^2 \wt k_k  -(y\wt k) v_k\Bigr) \Biggr]
+  a_i{}^h{}_{hk}
+  \theta_i.
\ee

{

\vskip 1cm

We find
$$
 \theta_i=
a^{nm}\Biggl[
a_{nikm}  +  \fr g{B}   \Bigl(q(b+gq)b_i+qv_i\Bigr) b_ja_n{}^j{}_{km}
+\fr g{B}\Bigl (qb_i-b\fr{v_i}q\Bigr) v_j a_n{}^j{}_{km}
\Biggr]
$$

\ses

$$
+\fr gq  bb^nb^m\Biggl[
a_{nikm}  +  \fr g{B}   \Bigl(q(b+gq)b_i+qv_i\Bigr) b_ja_n{}^j{}_{km}
+\fr g{B}\Bigl (qb_i-b\fr{v_i}q\Bigr) v_j a_n{}^j{}_{km}
\Biggr]
$$

\ses

$$
-\fr gq  b^ny^m\Biggl[
a_{nikm}  +  \fr g{B}   \Bigl(q(b+gq)b_i+qv_i\Bigr) b_ja_n{}^j{}_{km}
+\fr g{B}\Bigl (qb_i-b\fr{v_i}q\Bigr) v_j a_n{}^j{}_{km}
\Biggr]
$$

\ses

$$
-\fr gq  y^nb^m\Biggl[
a_{nikm}  +  \fr g{B}   \Bigl(q(b+gq)b_i+qv_i\Bigr) b_ja_n{}^j{}_{km}
+\fr g{B}\Bigl (qb_i-b\fr{v_i}q\Bigr) v_j a_n{}^j{}_{km}
\Biggr]
$$

\ses

$$
+\fr g{Bq}(b+gq)y^my^n   \Biggl[
a_{nikm}  +  \fr g{B}   \Bigl(q(b+gq)b_i+qv_i\Bigr) b_ja_n{}^j{}_{km}
+\fr g{B}\Bigl (qb_i-b\fr{v_i}q\Bigr) v_j a_n{}^j{}_{km}
\Biggr].
$$

{

The previous representation reduces to
$$
 \theta_i=
a^{nm}\Biggl[
a_{nikm}  +  \fr g{Bq}   \Bigl(gq^3b_i+(q^2+b^2)v_i\Bigr) b_ja_n{}^j{}_{km}
+\fr g{B}\Bigl (qb_i-b\fr{v_i}q\Bigr) y^j a_{njkm}
\Biggr]
$$

\ses

$$
+\fr gq  bb^nb^m\Biggl[
a_{nikm}
+\fr g{B}\Bigl (qb_i-b\fr{v_i}q\Bigr) y^j a_{njkm}
\Biggr]
$$

\ses

$$
-\fr gq  b^ny^m\Biggl[
a_{nikm}
+\fr g{B}\Bigl (qb_i-b\fr{v_i}q\Bigr) y^j a_{njkm}
\Biggr]
$$

\ses

$$
-\fr gq  y^nb^m\Biggl[
a_{nikm}    +  \fr g{Bq}   \Bigl(gq^3b_i+(q^2+b^2)v_i\Bigr) b_ja_n{}^j{}_{km}
\Biggr]
$$

\ses

$$
+\fr g{Bq}(b+gq)y^my^n   \Biggl[
a_{nikm}   +  \fr g{Bq}   \Bigl(gq^3b_i+(q^2+b^2)v_i\Bigr) b_ja_n{}^j{}_{km}
\Biggr],
$$

{

or
$$
 \theta_i=
  a^{nm}\Biggl[
a_{nikm}
+\fr g{B}\Bigl (qb_i-b\fr{v_i}q\Bigr) y^j a_{njkm}
\Biggr]
+ \fr g{Bq}(b+gq)y^my^n   a_{nikm}
$$

\ses

$$
+      \fr g{Bq}   \Bigl(gq^3b_i+(q^2+b^2)v_i\Bigr) b_ja_n{}^j{}_{km} a^{nm}
$$

\ses

$$
+\fr gq  bb^nb^m\Biggl[
a_{nikm}
+\fr g{B}\Bigl (qb_i-b\fr{v_i}q\Bigr) y^j a_{njkm}
\Biggr]
$$

\ses

$$
-\fr gq  b^ny^m\Biggl[
a_{nikm}
+\fr g{B}\Bigl (qb_i-b\fr{v_i}q\Bigr) y^j a_{njkm}
\Biggr]
$$

\ses

$$
-\fr gq  y^nb^m\Biggl[
a_{nikm}    +  \fr g{Bq}   \Bigl(gq^3b_i+(q^2+b^2)v_i\Bigr) b_ja_n{}^j{}_{km}
\Biggr]
$$

\ses

\be
+\fr g{Bq}(b+gq)
 \fr g{Bq}   \Bigl(gq^3b_i+(q^2+b^2)v_i\Bigr) b_ja_n{}^j{}_{km}
y^my^n.
\ee

{

Make required insertions
\be
b_ja_n{}^j{}_{km}=\Bigl(-\wt k_kr_{nm}+\wt k_mr_{nk}\Bigr),
\qquad
b^na_n{}^h{}_{hm}=\wt k_hr^h{}_m  -(N-1)\wt k_m
\ee
(see (A.48))
and   obtain
$$
 \theta_i   =
 a^{nm}\Biggl[
a_{nikm}
+\fr g{B}\Bigl (qb_i-b\fr{v_i}q\Bigr) y^j a_{njkm}
\Biggr]
+ \fr g{Bq}(b+gq)y^my^n   a_{nikm}
$$

\ses

$$
- \fr g{Bq}   \Bigl(gq^3b_i+(q^2+b^2)v_i\Bigr) (N-1) \wt k_k  + \fr g{Bq} \Bigl(gq^3b_i+(q^2+b^2)v_i\Bigr) r_k{}^m\wt k_m
$$

\ses

$$
+\fr gq  bb^m\Biggl[
\wt k_kr_{im}-\wt k_mr_{ik}
+\fr g{B}\Bigl (qb_i-b\fr{v_i}q\Bigr) y^j \Bigl(\wt k_kr_{jm}-\wt k_mr_{jk}\Bigr)
\Biggr]
$$

\ses

$$
-\fr gq  y^m\Biggl[
\wt k_kr_{im}-\wt k_mr_{ik}
+\fr g{B}\Bigl (qb_i-b\fr{v_i}q\Bigr)y^j \Bigl(\wt k_kr_{jm}-\wt k_mr_{jk}\Bigr)
\Biggr]
$$

\ses

$$
-\fr gq  y^n\Biggl[
-\wt k_nr_{ki}+\wt k_ir_{nk}    +  \fr g{Bq}   \Bigl(gq^3b_i+(q^2+b^2)v_i\Bigr)
\Bigl(-\wt k_kr_{nm}+\wt k_mr_{nk}\Bigr)   b^m
\Biggr]
$$

\ses

\be
-\fr g{Bq}(b+gq)
 \fr g{Bq}   \Bigl(gq^3b_i+(q^2+b^2)v_i\Bigr)
\Bigl(q^2 \wt k_k  -(y\wt k) v_k\Bigr).
\ee

{

We obtain the tensor
$$
R^m{}_{ikm}         +   R_i{}^h{}_{hk}
=\fr 12 k^2g^2 (N-2)r_{ik}
+ \fr {g^2} {2}
 b^m\wt k_m  \lf( r_{ik}-\fr1{q^2}v_iv_k\rg)
$$

\ses    \ses

$$
+\fr g{B}\Bigl (qb_i-b\fr{v_i}q\Bigr)
\Biggl[
\fr 14 k^2g^2 (N-2)v_k
+ \fr {gq} {2}
\Biggl( \wt k_m r_k^m   -  \fr1{q^2} \wt k_mv^m v_k   - N \wt k_k \Biggr)
\Biggr]
$$

\ses

\ses

$$
+\fr g{Bq}(b+gq)
\Biggl[
\fr 14 k^2g^2q^2  \lf( r_{ik}-\fr1{q^2}v_iv_k\rg)
- \fr12 gq
(y \wt k)  \lf( r_{ik}-\fr1{q^2}v_iv_k\rg)
$$

\ses

\ses

$$
+ \fr g {q}
 v_i\Bigl(q^2 \wt k_k  - (y\wt k) v_k\Bigr)
\Biggr]
+  a_i{}^h{}_{hk}
$$

    \ses  \ses

$$
+
 a^{nm}\Biggl[
a_{nikm}
+\fr g{B}\Bigl (qb_i-b\fr{v_i}q\Bigr) y^j a_{njkm}
\Biggr]
+ \fr g{Bq}(b+gq)y^my^n   a_{nikm}
$$

\ses\ses

$$
- \fr g{Bq}   \Bigl(gq^3b_i+(q^2+b^2)v_i\Bigr)(N-1) \wt k_k  + \fr g{Bq} \Bigl(gq^3b_i+(q^2+b^2)v_i\Bigr) r_k{}^m\wt k_m
$$

\ses\ses

$$
+\fr gq  bb^m\Biggl[
-\wt k_mr_{ik}
-\fr g{B}\Bigl (qb_i-b\fr{v_i}q\Bigr) y^j \wt k_mr_{jk}
\Biggr]
$$

\ses\ses

$$
-\fr gq  y^m\Biggl[
\wt k_kr_{im}-\wt k_mr_{ik}
+\fr g{B}\Bigl (qb_i-b\fr{v_i}q\Bigr)y^j \Bigl(\wt k_kr_{jm}-\wt k_mr_{jk}\Bigr)
\Biggr]
$$

\ses\ses

$$
-\fr gq  y^n\Biggl[
-\wt k_nr_{ki}+\wt k_ir_{nk}    +  \fr g{Bq}   \Bigl(gq^3b_i+(q^2+b^2)v_i\Bigr)
\wt k_mr_{nk}  b^m
\Biggr]
$$

\ses\ses

\be
-\fr g{Bq}(b+gq)
 \fr g{q}v_i
\Bigl(q^2 \wt k_k  -(y\wt k) v_k\Bigr),
\ee

{

which can be simplified to read        
\ses
$$
R^m{}_{ikm}         +   R_i{}^h{}_{hk}
=\fr 12 k^2g^2 (N-2)r_{ik}
+ \fr {g^2} {2}
 b^m\wt k_m  \lf( r_{ik}-\fr1{q^2}v_iv_k\rg)
$$

\ses    \ses

$$
+\fr g{B}\Bigl (qb_i-b\fr{v_i}q\Bigr)
\Biggl[
\fr 14 k^2g^2 (N-2)v_k
+ \fr {gq} {2}
\Biggl( \wt k_m r_k^m   -  \fr1{q^2} \wt k_mv^m v_k   - N \wt k_k \Biggr)
\Biggr]
$$

\ses

\ses

$$
+\fr g{Bq}(b+gq)
\Biggl[
\fr 14 k^2g^2q^2  \lf( r_{ik}-\fr1{q^2}v_iv_k\rg)
- \fr12 gq
(y \wt k)  \lf( r_{ik}-\fr1{q^2}v_iv_k\rg)
\Biggr]
$$

              \ses  \ses

$$
+ 2 a_i{}^h{}_{hk}
+\fr g{B}\Bigl (qb_i-b\fr{v_i}q\Bigr) y^j a_{njkm}  a^{nm}
+ \fr g{Bq}(b+gq)y^my^n   a_{nikm}
$$

\ses\ses

$$
- \fr g{Bq}   \Bigl(gq^3b_i+(q^2+b^2)v_i\Bigr) (N-1) \wt k_k  + \fr g{Bq} \Bigl(gq^3b_i+(q^2+b^2)v_i\Bigr) r_k{}^m\wt k_m
$$

\ses\ses

$$
+\fr gq  bb^m\Biggl[
-\wt k_mr_{ik}
-\fr g{B}\Bigl (qb_i-b\fr{v_i}q\Bigr) \wt k_m v_k
\Biggr]
$$

\ses\ses

$$
-\fr gq  \Biggl[
\wt k_kv_i-(y\wt k)   r_{ik}
+\fr g{B}\Bigl (qb_i-b\fr{v_i}q\Bigr)     \Bigl(q^2 \wt k_k  -(y\wt k) v_k\Bigr)
\Biggr]
$$

\ses\ses

\be
-\fr gq  \Biggl[
-(y\wt k)r_{ki}+\wt k_iv_k +\fr gqv_i\wt k_mb^m v_k     +  \fr {g^2}{B}   \Bigl(q^2b_i-bv_i\Bigr)
\wt k_mb^m v_k
\Biggr].
\ee

\ses\ses

Thus, the following proposition is valid.

\ses\ses

{\large Proposition B2.} {\it  In any Landsberg case of
the $\cF\cF^{PD}_g $--space,
the covariantly conserved  tensor
$$
\rho_{ij} =\fr12(R_i{}^m{}_{mj}+R^m{}_{ijm})-\fr12g_{ij}R^{mn}{}_{nm}
$$
is given by the explicit representations
}
(B.53)  {\it  and}  (B.39).

\ses\ses

{

Contracting yields

$$
y^iy^k\Bigl(R^m{}_{ikm}         +   R_i{}^h{}_{hk}\Bigr)
=\fr 12 k^2g^2 (N-2)q^2
+ 2 a_i{}^h{}_{hk}y^iy^k
$$

\ses

\be
-  gq  (N-1) (y\wt k)
  +  gq(v^m\wt k_m)
- gbq(b\wt k)
-\fr gq   gq^3\wt k_mb^m,
\ee

{


as well as
$$
y^k\Bigl(R^m{}_{ikm}         +   R_i{}^h{}_{hk}\Bigr)
=\fr 12 k^2g^2 (N-2)v_i
$$

\ses    \ses

$$
+\fr g{B}\Bigl (qb_i-b\fr{v_i}q\Bigr)
\Biggl[
\fr 14 k^2g^2 (N-2)q^2
- \fr {gq} {2}
 N (y\wt k)
\Biggr]
$$

\ses

\ses

          \ses

$$
+ 2 a_i{}^h{}_{hk}y^k
+\fr g{B}\Bigl (qb_i-b\fr{v_i}q\Bigr) y^j y^k a_{njkm}  a^{nm}
$$

\ses

$$
- (N-1) \fr g{Bq}   \Bigl(gq^3b_i+(q^2+b^2)v_i\Bigr)  (y\wt k)
  + \fr g{Bq} \Bigl(gq^3b_i+(q^2+b^2)v_i\Bigr)(v\wt k)
$$

\ses

$$
+\fr gq  b\Biggl[
- v_i   -\fr g{B}\Bigl (qb_i-b\fr{v_i}q\Bigr)  q^2
\Biggr]
(b\wt k)$$

\ses

\ses

\be
-\fr gq  \Biggl[
-(y\wt k)v_i+\wt k_iq^2 + gqv_i (b\wt k)      +  \fr {g^2q^2}{B}   \Bigl(q^2b_i-bv_i\Bigr) (b\wt k)
\Biggr].
\ee

\ses

{

\vskip 1cm

\setcounter{equation}{0}

\nin
{\bf \large Appendix C: ~  Special--case evaluation}

\ses\ses

In the special case characterized by the warped representation (1.24) of the interval   square
$(ds)^2$
the  vector $\wt k_m$ is a factor of the fundamental vector
$b_m$:
\be
\wt k_m=fb_m, \qquad   f=(b\wt k)
\ee
(see (1.19)), which  entails
\be
\wt k_mr^m{}_n=0,
\ee
\be
 \wt k_mv^m=0,
\ee
and
\be
 n_1=0
\ee
(the function $n_1$ was defined in (B.15)).

{

Also, it follows that
\be
\phi=\phi(z^N),
\ee
so  that
\be
k =k(z^N).
\ee
Let us put
\be
\dot k=\fr{dk}{dz^N}, \qquad   \dot {\phi}=\fr{d \phi}{dz^N}
\ee
and obtain
\be
(b\wt k)= \dot k + k^2.
\ee

When proceeding in this way, we have
\be
r_{ab}(z^A)=\lf(\phi(z^N)\rg)^2p_{ab}(z^c)
\ee
and
\be
(ds)^2=\lf(dz^N\rg)^2+\lf(\phi(z^N)\rg)^2p_{ab}(z^c)dz^adz^b,
\ee
together with
\be
 \dot {\phi} =k\phi
\ee
and
$$
\fr1{\phi} \ddot\phi=\dot k+k^2.
$$

\ses

The Christoffel symbols (A.111) reduce to
\be
s^a{}_{bN}= k\de^a_b,          \qquad s^N{}_{bc}=-  k r_{bc},
\qquad
 s^a{}_{bc}=\fr12p^{ad}\Biggl(\D{p_{dc}}{z^b}+\D{p_{bd}}{z^c}-\D{p_{bc}}{z^d}\Biggr).
\ee

{

If we calculate from them the Riemannian curvature tensor
(see  (A.44) in Appendix A), we obtain the components
\be
a_a{}^N{}_{bN}=\D{s^N{}_{aN}}{z^b}-\D{s^N{}_{ab}}{z^N}+  s^h{}_{aN}s^N{}_{hb}-  s^h{}_{ab}s^N{}_{hN}
=
\dot kr_{ab} -  k^2 r_{ab}
\ee
and
\be
a_a{}^N{}_{bc}=\D{s^N{}_{ac}}{z^b}-\D{s^N{}_{ab}}{z^c}+s^h{}_{ac}s^N{}_{hb}-s^h{}_{ab}s^N{}_{hc}
=
\D{s^N{}_{ac}}{z^b}-\D{s^N{}_{ab}}{z^c}+s^f{}_{ac}s^N{}_{fb}-s^f{}_{ab}s^N{}_{fc},
\ee
\ses
together with
\be
a_a{}^e{}_{bc}=\D{s^e{}_{ac}}{z^b}-\D{s^e{}_{ab}}{z^c}+s^h{}_{ac}s^e{}_{hb}-s^h{}_{ab}s^e{}_{hc}
=P_a{}^e{}_{bc}     +     s^N{}_{ac}s^e{}_{Nb}    -    s^N{}_{ab}s^e{}_{Nc}.
\ee

{

Under these conditions, we obtain
\be
a_a{}^N{}_{bN}=(\dot k  -  k^2) r_{ab},
\ee

\ses

\be
a_a{}^N{}_{bc}   =   0,
\ee

\ses

\be
a_a{}^e{}_{bc}=P_a{}^e{}_{bc}     -  k^2(r_{ac}\de^e_b    - r_{ab}\de^e_c),
\ee
\ses
where
\be
P_a{}^e{}_{bc}=\D{p^e{}_{ac}}{z^b}-\D{p^e{}_{ab}}{z^c}+p^f{}_{ac}p^e{}_{fb}-p^f{}_{ab}p^e{}_{fc}.
\ee
\ses
We have
\be
a_N{}^h{}_{Nh}=(N-1) (\dot k - k^2),
\ee
\ses
\be
a_a{}^h{}_{Nh}=0,
\ee
\ses
and
\be
a_a{}^h{}_{bh}=P_{ab}   +       (N-2)k^2r_{ab} + (\dot k - k^2)r_{ab}.
\ee

\ses

In terms of the quantities
\be
\xi_1 = - k^2-\fr1{q^2(N-2)}y^ay^bP_{ab},
\qquad
\xi_2= - k^2-\fr1{(N-1)(N-2)}a^{ab}P_{ab},
\ee
we obtain  the concise representation
\be
y^ny^ma_n{}^h{}_{mh}=
(N-1) (\dot k - k^2)b^2     -(N-2)\xi_1q^2 + (\dot k - k^2)q^2.
\ee
Notice that
\be
y^N=b.
\ee
Henceforth, we shall use the notation
\be
 \beta =\dot k-k^2 \equiv (b\wt k)-2k^2.
\ee

{

The representation
$$
a_n{}^i{}_{km}=r_n^h   r^i_l   r_k^j   r_m ^u  P_h{}^l{}_{ju}
-   k^2\bigl(r_{nm}r^i_k-  r_{nk}r^i_m \bigr )
$$

\ses

\be
+(b\wt k)  \lf(b^ib_mr_{nk}-b^ib_kr_{nm}   -  b_nb_mr^i_k+b_nb_kr^i_m\rg)
\ee
is valid, entailing
$$
a_n{}^h{}_{mh}=r_n^i  r_m^j  P_{ij}-(N-2)k^2r_{nm} + (b\wt k) r_{nm}+
(N-1)(b\wt k) b_nb_m.
$$

{

The scalar $\mu$ given by (B.41) reduces now to
$$
\mu=- 2 (N-1) \beta
+  (N-1) (N-2)\xi_2
$$

\ses

$$
- (N-1) \fr{g^2q^2} B  \beta
+  \fr {gq(b+gq) }   {B}(N-2)\Bigl(\xi_1  +   \beta\Bigr).
$$

{

The contraction (B.54) is got simplified to read

$$
y^iy^k      \Bigl(R^m{}_{ikm}         +   R_i{}^h{}_{hk}\Bigr)
=\fr 12 k^2g^2 (N-2)q^2
-2 (N-1)b^2  \beta
$$

\ses

\be
+2(N-2)\xi_1 q^2    -     2q^2  \beta
-  N gbq  (b\wt k)
- g^2q^2 (b\wt k),
\ee

{

\nin
and  (B.55) becomes
$$
y^k\Bigl(R^m{}_{ikm}         +   R_i{}^h{}_{hk}\Bigr)
=
\fr 12 k^2g^2 (N-2)v_i
$$

\ses

\ses

$$
 +\fr {g^2}{B}\Bigl (qb_i-b\fr{v_i}q    \Bigr)
\Biggl[  \fr 14 k^2g (N-2)q^2  - \fr {bq} {2}   N (b\wt k)  \Biggr]
+ 2 a_i{}^h{}_{hk}y^k
$$

\ses

\ses

$$
+\fr g{B}\Bigl (qb_i-b\fr{v_i}q\Bigr) y^j y^k a_{njkm}  a^{nm}
$$

\ses   \ses

$$
- (N-1) \fr{ gb}{Bq}   \Bigl(gq^3b_i+(q^2+b^2)v_i\Bigr)  (b\wt k)
$$

\ses  \ses

\be
-gq\wt k_i
 - g^2v_i (b\wt k)      -  \fr {g^2(b+gq)}{B}   \Bigl(q^2b_i-bv_i\Bigr) (b\wt k).
\ee

{

The scalar  (B.39) reduces to
$$
K^2 R^{ni}{}_{in}
= \fr14  (N-2)  (N-1) k^2g^2    B
+ \fr14  (N-2)   \fr1b k^2g^3 q(b^2+gbq + q^2)
$$

\ses

$$
-  \fr {gq}{b}
 \Biggl[
 \fr 14(N-2)  k^2g^2 q^2
-\fr 12  gb q
 N (b\wt k)
 \Biggr]
 $$

\ses

$$
- 2 (N-1)\beta B
 +  (N-1) (N-2)\xi_2 B
$$

\ses

\be
- (N-1) g^2q^2  \beta
+  (N-2) gq(b+gq) \Bigl(\xi_1+   \beta \Bigr),
\ee
where
$\xi_1,\,\xi_2$ are the quantities which were defined in (C.23).

{

Simplifying yields
$$
K^2 R^{ni}{}_{in}
=
- 2 (N-1)B \beta
 +  (N-1) (N-2)\lf(\xi_2 +\fr14g^2k^2\rg) B
+\fr12 (N-2) g^2q^2  (b\wt k)
$$

\ses

\be
+  (N-2) gq(b+gq)\lf(\xi_1 +\fr14g^2k^2\rg)
+  (N-2) gbq   \beta
+2g^2q^2k^2.
\ee

{

Henceforth,  we restrict the treatment to the
{\it constant--curvature case}:
\be
P_a{}^e{}_{bc} =  - \ka_1 (p_{ac}\de^e_b    - p_{ab}\de^e_c), \qquad \ka_1=-1, 0, 1,
\ee
which can be written as
\be
P_a{}^e{}_{bc} =  - \ka (r_{ac}\de^e_b    - r_{ab}\de^e_c)
\ee
with
\be
\ka=\fr1{\phi^2}\ka_1
\ee
(see (C.9)).
The  contracted tensor
\be
P_{ab}    =   P_a{}^e{}_{be}
\ee
equals
\be
P_{ab}=(N-2)\ka r_{ab},
\ee
entailing
\be
y^ay^bP_{ab}   =(N-2)\ka q^2, \qquad
a^{ab} P_{ab}  = (N-1)(N-2)\ka.
\ee
The quantities (C.23) become
\be
\xi_1=\xi_2=       -      k^2-\ka   \equiv \xi.
\ee
We obtain
\be
a_a{}^h{}_{bh}=-(N-2)\xi r_{ab} +  \beta r_{ab}
\ee
\ses
and
\be
y^ny^ma_n{}^h{}_{mh}=
(N-1)  \beta b^2     -(N-2)\xi q^2 +  \beta  q^2,
\ee
together with
\be
a_i{}^h{}_{hk}=     -(N-1)   \beta b_ib_k+ (N-2)\xi r_{ik} -  \beta r_{ik},
\ee

\ses\ses

\be
-y^j a_j{}^h{}_{hk}=         (N-1)  \beta  bb_k-(N-2)\xi v_k +  \beta v_k,
\ee
\ses
and
\be
y^my^n   a_{nikm}          =  - \beta\Bigl[b^2r_{ik}+q^2b_ib_k-bb_iv_k-bb_kv_i\Bigr]
  +\xi (q^2r_{ik}-v_iv_k).
\ee

\ses

\ses\ses

{

With (C.36)--(C.42),  the tensor (C.29) reduces to
$$
y^k\Bigl(R^m{}_{ikm}         +   R_i{}^h{}_{hk}\Bigr)=
\fr 12 k^2g^2 (N-2)v_i
$$

\ses\ses

$$
-\fr {g^2q}{B}
\Biggl[
 \fr 14 k^2g (N-2)q^2  - \fr {bq} {2}   N (b\wt k)\Biggr]   e_i
- 2(N-1) \beta b b_i
 +      2(N-2)\xi  v_i    -     2 \beta v_i
$$

\ses   \ses

$$
-(N-2)  \fr {gq^3}{B}    \xi         e_i
+
\fr {gq}{B}
\Bigl[
 (N-1)b^2 +q^2
 \Bigr]  \beta
e_i
$$

\ses    \ses

$$
- (N-1) \fr{ gq}{B}   \Bigl(Bb_i  +(q^2+b^2)e_i  \Bigr)  (b\wt k)
$$

\ses  \ses

$$
-gq\wt k_i
 - g^2v_i (b\wt k)     +  \fr {g^2q^2(b+gq)}{B}   (b\wt k)e_i
$$

\ses\ses           \ses\ses

$$
=
\fr {g^2q}{B}
\fr {bq} {2}   N (b\wt k)   e_i
- 2(N-1)  \beta b b_i
 +      2(N-2)\xi_9  v_i    -     2 \beta v_i
$$

\ses   \ses

$$
-(N-2)  \fr {gq^3}{B}    \xi_9         e_i
+
\fr {gq}{B}
\Bigl[
- (N-2)q^2
+ gq(b+gq)
 \Bigr] (b\wt k)
e_i
$$

\ses    \ses

$$
- Ngqb_i  (b\wt k)
  - g^2v_i (b\wt k)
-2(N-1)\fr{gqb^2}B k^2e_i
-2\fr{gq^3}Bk^2e_i
$$

{


$$
=
\fr {g^2q}{B}
 \fr {bq} {2}   N (b\wt k) e_i
-(N-2)  \fr {gq^3}{B}    \xi_9  e_i
+
\fr {gq}{B}
\Bigl[
- (N-2)q^2
+ gq(b+gq)
 \Bigr] (b\wt k)
e_i
$$

\ses    \ses

$$
 +      2(N-2)\xi_9  \fr{q^2}be_i    -     2 \beta \fr{q^2}be_i
  - g^2 (b\wt k)\fr{q^2}be_i
- 2(N-1)  \beta b b_i
- Ngq (b\wt k)  b_i
$$

\ses    \ses

$$
 +      2(N-2)\xi_9  \fr{q^2}bb_i    -     2  \beta \fr{q^2}bb_i
  - g^2 (b\wt k)\fr{q^2}bb_i
-2(N-1)\fr{gqb^2}B k^2e_i
-2\fr{gq^3}Bk^2e_i,
$$
where
$$
\xi_9=\xi+\fr14g^2k^2,
$$
or, by virtue of (C.38),
\be
\xi_9=-\lf(1-\fr{g^2}4\rg)k^2   -\ka.
\ee
We have used the vector
\be
e_i=-b_i+\fr b{q^2}v_i.
\ee
If we remind
$$
y_i=\Bigl(v_i+(b+gq)b_i\Bigr)\fr{K^2}B
$$
(see (A.23)), we may write the equality
$$
y_i=(q^2e_i+Bb_i)  \fr{K^2}{bB}.
$$

{

This yields
$$
y^k\Bigl(R^m{}_{ikm}         +   R_i{}^h{}_{hk}\Bigr)=
\fr {g^2q}{B}
\fr {bq} {2}   N (b\wt k)   e_i
$$

\ses\ses

$$
-(N-2)  \fr {gq^3}{B}    \xi_9         e_i
-(N-2) \fr {gq^3}{B}  (b\wt k)  e_i
- \fr {g^2q^4}{Bb}  (b\wt k)  e_i
$$

\ses    \ses

$$
 +      2(N-2)\xi_9  \fr{q^2}be_i    -     2(b\wt k) \fr{q^2(b^2+gbq+q^2)}{Bb}e_i
$$

\ses

\ses

$$
+ 2(N-1)(b\wt k) b \fr{q^2}B e_i
+ Ngq (b\wt k) \fr{q^2}B  e_i
$$

\ses    \ses

$$
-      2(N-2)\xi_9  \fr{q^2}b  \fr{q^2}B  e_i    +     2  \beta \fr{q^2}b  \fr{q^2}B  e_i
+ g^2 (b\wt k)\fr{q^2}b  \fr{q^2}B  e_i
+R_9\fr1{K^2}y_i
$$

\ses\ses

$$
-2(N-1)\fr{gqb^2}B k^2e_i
-2\fr{gq^3}Bk^2e_i,
$$
\ses
where
$$
R_9=
- 2(N-1) \beta  b^2
- Ngq (b\wt k)  b
 +      2(N-2)\xi_9  q^2  -     2 \beta q^2
  - g^2 (b\wt k)q^2,
$$
\ses
or
$$
R_9=
- 2(N-2) \beta b^2
- (N-2)gq (b\wt k)  b
 +      2(N-2)\xi_9  q^2  -     2(b\wt k) B
  - g^2 (b\wt k)q^2
  +4(b^2+q^2)k^2.
$$

\ses

{

In this way we arrive at the equality
$$
y^k\Bigl(R^m{}_{ikm}         +   R_i{}^h{}_{hk}\Bigr)=
(N-2)I_4  q  e_i
+ N \fr {g^2bq^2}{2B}  (b\wt k)e_i
+4\fr{gq^3}B k^2e_i
$$

\ses\ses

\be
-2(N-1)\fr{gqb^2}B k^2e_i
-2\fr{gq^3}Bk^2e_i
  +R_9   \fr1{K^2}y_i
\ee
with          
$$
I_4=
- \fr {gq^2}{B}    \xi_9
 +      2\xi_9  \fr{q}b
+ 2(b\wt k) \fr{bq}B
-    2\xi_9   \fr{q^3}{Bb},
$$
or
\be
I_4=
 \fr {gq^2}{B}    \xi_9
+ 2\Bigl(\xi_9 +(b\wt k)\Bigr) \fr{bq}B.
\ee

\ses

{

The scalar (C.31) takes on the form
$$
K^2 R^{ni}{}_{in}
=
- 2 (N-1) \beta B
 +  (N-1) (N-2) \xi_9B
$$

\ses

\be
+\fr12 (N-2) g^2q^2 (b\wt k)
+  (N-2) gq(b+gq) \xi_9
+  (N-2) gbq  \beta
+2g^2q^2k^2.
\ee

{

Now we calculate the quantity
\be
 \epsilon =   \fr12(  R_9  -   K^2 R^{ni}{}_{in}).
 \ee
With (C.48), we find
\be
R_9-   K^2 R^{ni}{}_{in}=(N-2)E_4  -g^2q^2(b\wt k)
-2k^2B+4q^2k^2-2g^2q^2k^2,
\ee
where
\be
E_4=
 2\lf(1-\fr{g^2}4\rg)   (b\wt k)q^2  +      2\xi_9  q^2   -  (N-1) \xi_9  B  -    gq(b+gq) \xi_9
 -4q^2k^2       -  2gbqk^2.
\ee
Thus we get
$$
\epsilon = \fr12 (N-2)E_4  - \fr12 g^2q^2(b\wt k)
-2k^2B+2(b^2+q^2) k^2-g^2q^2k^2,
$$
\ses
or
\be
\epsilon = \fr12 (N-2)E_4  - \fr12 g^2q^2(b\wt k)
-2gbqk^2-g^2q^2k^2.
\ee

{


Now we are able to evaluate the explicit expression of the vector
$$
\rho_i~:=\rho_{ij}y^j.
$$
From (A.65) we get
\be
2\rho_i =y^k\Bigl(R^m{}_{ikm}         +   R_i{}^h{}_{hk}\Bigr)-y_iR^{mn}{}_{nm}.
\ee
We obtain
$$
2\rho_i=  (N-2)I_4qe_i   +  \fr {g^2bq^2}{2B} N (b\wt k)e_i
+4\fr{gq^3}B k^2e_i
$$

\ses

\be
-2(N-1)\fr{gqb^2}B k^2e_i
-2\fr{gq^3}Bk^2e_i
+\epsilon  \fr1{K^2}y_i.
\ee
If we apply here the formula (A.36) which introduces the vector $A_i$, we can obtain
 the expansion
\be
\rho_i= \epsilon \fr1{K^2} y_i  -r_1qA_i,
\ee
where
\be
r_1  = \fr{ b}{2 g K}\Biggl[
\fr{2 (N-2)}{N}P_4   +  g^2  (b\wt k)
+\fr{4gq}{Nb}k^2
-\fr{4(N-1)}Ng\fr bq k^2
\Biggr]
\ee
\ses
with
\be
P_4=
 g\fr qb  \xi_9
+ 2\Bigl(\xi_9 +(b\wt k)\Bigr)
\equiv \fr BqI_4.
\ee

{

Raising the index yields the expansion
\be
\rho^i= \epsilon  \fr1{K^2} y^i  -r_1qA^i.
\ee
We may apply here the representation (A.37) of the vector $A^i$, obtaining
\be
\rho^i= \epsilon  \fr1{K^2} y^i  - s_1\fr{1}  {K^2}\Bigl[Bb^i-(b+gq)y^i\Bigr],
\ee
where
\be
s_1  = \fr{ Nb}{4 }\Biggl[
\fr{2 (N-2)}{N}P_4   +  g^2  (b\wt k)
+\fr{8gq}{Nb}k^2
\Biggr].
\ee
In another convenient form,
\be
\rho^i=  - s_1\fr{B}  {K^2}b^i + \fr P {K^2} y^i,
\ee
where
\be
P=  \epsilon + s_1(b+gq).
\ee

{

Let us apply the {\it osculation procedure}, specifying the variable $y^h$ according to
\be
y^h=b^h(x).
\ee
Noting that
$$
q_{|_{y^h=b^h}}=0,  \qquad     b_{|_{y^h=b^h}}=1,
$$
from (C.52) we get
\ses
\be
\epsilon_{|_{y^h=b^h}} =      -  \fr12  (N-1)   (N-2)  \xi_9
\ee
and (C.60) yields us
\be
s_{1|_{y^h=b^h}}  = (N-2)\xi_9+
\lf(N-2+ \fr{ N}{4 }g^2\rg)
  (b\wt k).
  \ee
We have
\be
\rho^i_{|_{y^h=b^h}}  = \epsilon_{|_{y^h=b^h}} b^i
\ee
together with
\be
P_{|_{y^h=b^h}} = \cP,
\ee
where
$$
\cP=\epsilon_{|_{y^h=b^h}}+    s_{1|_{y^h=b^h}}
$$
(see (C.62)),
that is,
\be
\cP=-\fr12 (N-2) (N-3)  \xi_9     + \Bigl(N-2   + \fr{ N}{4 } g^2 \Bigr) (b\wt k)      .
\ee
\ses

{

From  (C.61) we infer
$$
 \lf(\D{\bigl(\fr{\rho^i}b\bigr)}{y^i}\rg)_{|_{y^h=b^h}}=
\lf(\D{\Bigl(\fr {P y^i} {bK^2}\Bigr)}{y^i}\rg)_{|_{y^h=b^h}}
=-\lf(\fr P {bK^2} \rg)_{|_{y^h=b^h}}
+N\lf(\fr P {bK^2} \rg)_{|_{y^h=b^h}}
=(N-1)\cP,
 $$
that is,
\be
 \lf(\D{\bigl(\rho^i/b\bigr)}{y^i}\rg)_{|_{y^h=b^h}}=   (N-1)\cP.
\ee
All the conditions which have underlined the equality (A.108) of Appendix A are fulfilled,
so that we are entitled to write
\be
 \nabla_i \rho^i_{\{b\}} =   (N-1)k\cP,
\ee
or, if we apply (A.109),
\be
b^i\partial_i \ga_{\{b\}}+(N-1) k  \ga_{\{b\}}    =    (N-1)k\cP.
\ee

From (C.66) and (A.105) it follows that
\be
 \ga_{\{b\}}=   \epsilon_{|_{y^h=b^h}}.
 \ee

{

\vskip 1cm

\setcounter{equation}{0}

\nin
{\bf \large Appendix D:~  Evaluation of the covariantly conserved tensor}       $\rho^i{}_k$

\ses\ses

Let us  evaluate the sum tensor
$R^m{}_{ikm}  +   R_i{}^h{}_{hk}$,
assuming  the special--case relations (C.1)--(C.12) to be valid.

\ses

From (B.53) it follows that
\ses\\
$$
R^m{}_{ikm}         +   R_i{}^h{}_{hk}
=\fr 12 k^2g^2 (N-2)r_{ik}
+ \fr {g^2} {2}
  \lf( r_{ik}-\fr1{q^2}v_iv_k\rg)
 (b\wt k)
 $$

\ses    \ses

$$
-\fr {gq}{B}
\Biggl[
\fr 14 k^2g^2 (N-2)v_k
- \fr {gq} {2}
 N \wt k_k
\Biggr]
e_i
$$

\ses

\ses

$$
+\fr g{Bq}(b+gq)
\Biggl[
\fr 14 k^2g^2q^2  \lf( r_{ik}-\fr1{q^2}v_iv_k\rg)
- \fr12 gbq
(b \wt k)  \lf( r_{ik}-\fr1{q^2}v_iv_k\rg)
\Biggr]
$$

              \ses  \ses

$$
+ 2 a_i{}^h{}_{hk}
-\fr {gq}{B}y^j a_j{}^h{}_{hk}
e_i
+ \fr g{Bq}(b+gq)y^my^n   a_{nikm}
$$

\ses\ses

$$
- (N-1)\fr g{Bq}   \Bigl(gq^3b_i+(q^2+b^2)v_i\Bigr)  \wt k_k
$$

\ses\ses

$$
+\fr gq  b\Biggl[
-r_{ik}
+\fr{ gq}{B}e_i  v_k
\Biggr]
(b\wt k)
+\fr{2gb}q(b\wt k)r_{ik}
-\fr gq  \Biggl[
\wt k_kv_i
-\fr {gq}{B}e_i   \Bigl(q^2 \wt k_k  -(y\wt k) v_k\Bigr)
\Biggr]
$$

\ses\ses

$$
-\fr gq  \Biggl[
\wt k_iv_k +\fr gqv_i(b\wt k) v_k     -  \fr {g^2q^2}{B} e_i
(b\wt k)
 v_k
\Biggr].
$$

{

Cancel similar terms and make insertion of the Riemannian curvature tensor contractions (C.41)--(C.43)
related to the isotropic case.
We obtain
$$
R^m{}_{ikm}         +   R_i{}^h{}_{hk}
=\fr 12 k^2g^2 (N-2)r_{ik}
+ \fr {g^2} {2}
  \lf( r_{ik}-\fr1{q^2}v_iv_k\rg)
 (b\wt k)
 $$

\ses    \ses

$$
-\fr {gq}{B}
\Biggl[
\fr 14 k^2g^2 (N-2)v_k
- \fr {gq} {2}
 N \wt k_k
\Biggr]
e_i
$$

\ses

\ses

$$
+\fr g{Bq}(b+gq)
\Biggl[
\fr 14 k^2g^2q^2  \lf( r_{ik}-\fr1{q^2}v_iv_k\rg)
- \fr12 gbq
(b \wt k)  \lf( r_{ik}-\fr1{q^2}v_iv_k\rg)
\Biggr]
$$

    \ses  \ses

$$
-2(N-1)\beta b_ib_k+ 2(N-2)\xi r_{ik} - 2\beta  r_{ik}
$$

\ses\ses

$$
+\fr {gq}{B}\Biggl((N-1)\beta bb_k-(N-2)\xi v_k +\beta  v_k\Biggr)
e_i
$$

\ses\ses

$$
+ \fr g{Bq}(b+gq)\Biggl[
 -\beta \Bigl[b^2r_{ik}+q^2b_ib_k-bb_iv_k-bb_kv_i\Bigr]
+\xi (q^2r_{ik}-v_iv_k)
\Biggr]
$$

\ses\ses

$$
- (N-1)\fr g{Bq}   \Bigl(gq^3b_i+(q^2+b^2)v_i\Bigr)  \wt k_k
$$

\ses\ses

$$
+\fr{gb}q(b\wt k)r_{ik}
-\fr gq  \Biggl[
v_i
-\fr {gq^3}{B}e_i
\Biggr]
\wt k_k
$$

\ses\ses

$$
-\fr gq  \Biggl[
\wt k_iv_k +\fr gqv_i(b\wt k) v_k     -  \fr {g^2q^2}{B} e_i
(b\wt k)
 v_k
\Biggr],
$$
where $\beta $ is the quantity (C.26).

{

\nin
 Here, the variable $\xi$ can conveniently be replaced by the variable
 $ \xi_9=\xi+\fr14g^2k^2$ (see (C.44)). Following in this way, we get
$$
R^m{}_{ikm}         +   R_i{}^h{}_{hk}
= \fr {g^2} {2}
  \lf( r_{ik}-\fr1{q^2}v_iv_k\rg)
 (b\wt k)
+\fr {gq}{B}
 \fr {gq} {2}
 N \wt k_k
e_i
$$

\ses

\ses

$$
-\fr12
\fr {g^2}{B}(B-q^2)
(b \wt k)  \lf( r_{ik}-\fr1{q^2}v_iv_k\rg)
$$

              \ses  \ses

$$
-2(N-1)\beta  b_ib_k+ 2(N-2)\xi_9   r_{ik} - 2\beta  r_{ik}
$$

\ses\ses

$$
+\fr {gq}{B}\Biggl((N-1)\beta  bb_k-(N-2)\xi_9   v_k +\beta  v_k\Biggr)
e_i
$$

\ses\ses

$$
- \fr g{Bq}(b+gq)
\Bigl[b^2r_{ik}+q^2b_ib_k-bb_iv_k-bb_kv_i\Bigr]
 \beta
 $$

\ses\ses

$$
+ \fr {gq}{B}(b+gq)
  \lf( r_{ik}-\fr1{q^2}v_iv_k\rg)
\xi_9
- (N-1)\fr g{Bq}   \Bigl(gq^3b_i+(q^2+b^2)v_i\Bigr)  \wt k_k
$$

\ses\ses

\be
+\fr{gb}q(b\wt k)r_{ik}
-\fr gq  \Biggl[
v_i
-\fr {gq^3}{B}e_i
\Biggr]
(b\wt k)b_k
-\fr gq  \Biggl[
b_i +\fr gqv_i   -  \fr {g^2q^2}{B} e_i
\Biggr]
(b\wt k)v_k.
\ee

{

Applying
$$
e_i=-b_i+\fr b{q^2}v_i
$$
leads to
$$
R^m{}_{ikm}         +   R_i{}^h{}_{hk}
=
\fr {gq}{B}
 \fr {gq} {2}
 N \wt k_k
e_i
+\fr12  \fr {g^2q^2}{B}
(b \wt k)  \lf( r_{ik}-\fr1{q^2}v_iv_k\rg)
$$

              \ses  \ses

$$
-2(N-1)\beta  b_ib_k+ 2(N-2)\xi_9   r_{ik} - 2\beta  r_{ik}
$$

\ses\ses

$$
+\fr {gq}{B}\Biggl((N-1) \beta  bb_k-(N-2)\xi_9   v_k + \beta  v_k\Biggr)
e_i
$$

\ses\ses

$$
- \fr g{qb}
\Bigl[b^2r_{ik}+q^2b_ib_k-bb_iv_k-bb_kv_i\Bigr]
\beta
 $$

\ses\ses

$$
+ \fr {gq}{Bb}
\Bigl[b^2r_{ik}+q^2b_ib_k-bb_iv_k-bb_kv_i\Bigr]
\beta
 $$

\ses\ses

$$
+ \fr {gq}{B}(b+gq)
  \lf( r_{ik}-\fr1{q^2}v_iv_k\rg)
\xi_9
$$

\ses\ses

$$
- (N-1)\fr g{Bq}  \fr{q^2}b B  (b \wt k) b_ib_k
- (N-1)\fr {gq}{Bb}(q^2+b^2)
 (b \wt k)b_ke_i
$$

\ses\ses

$$
+\fr{gb}q(b\wt k)r_{ik}
-\fr gq  \Biggl[    \fr{q^2}b b_i  +   \fr{q^2}b e_i
-\fr {gq^3}{B}e_i
\Biggr]
(b\wt k)b_k
$$

\ses\ses

$$
-\fr gq (b\wt k) b_iv_k
-\fr {g^2}{q^2}
(b\wt k) v_i v_k
+  \fr {g^3q}{B}
(b\wt k) e_i  v_k.
$$

{

\nin
Let us perform appropriate cancellation:
$$
R^m{}_{ikm}         +   R_i{}^h{}_{hk}
=
\fr {gq}{B}
 \fr {gq} {2}
 N (b\wt k)b_k
e_i
$$

\ses

\ses

$$
+\fr12  \fr {g^2q^2}{B}   (b \wt k)  \lf( r_{ik}-\fr1{q^2}v_iv_k\rg)
+        \fr {gq}{B}(b+gq)   \lf( r_{ik}-\fr1{q^2}v_iv_k\rg)     \xi_9
$$

              \ses  \ses

$$
-2(N-1)\beta  b_ib_k+ 2(N-2)\xi_9   r_{ik} - 2\beta  r_{ik}
$$

\ses\ses

$$
+\fr {gq}{B}\Biggl((N-1)\beta bb_k-(N-2)\xi_9   v_k +\beta  v_k\Biggr)
e_i
$$

\ses\ses

$$
- \fr g{qb}
\Bigl[q^2b_ib_k-bb_kv_i\Bigr]
\beta
 $$

\ses\ses

$$
+ \fr {gq}{Bb}
\Bigl[    b^2r_{ik}+      q^2b_ib_k-bb_iv_k-bb_kv_i\Bigr]
\beta
 $$

\ses\ses

$$
- N\fr {gq}b  (b \wt k) b_ib_k
- N\fr {gq}{Bb}(q^2+b^2)
 (b \wt k)b_ke_i
$$

\ses\ses

$$
-\fr {g^2}{q^2}
(b\wt k) v_i v_k
+  \fr {g^3q}{B}
(b\wt k) e_i  v_k
$$

{

$$
=
\fr N2 \fr {g^2q^2}{B}   (b\wt k)b_k   e_i
- N\fr {gq}{Bb}(q^2+b^2)    (b \wt k)b_ke_i$$

\ses

\ses

$$
+\fr12  \fr {g^2q^2}{B}  (b \wt k)  \lf( r_{ik}-\fr1{q^2}v_iv_k\rg)
+ \fr {gq}{B}(b+gq)    \lf( r_{ik}-\fr1{q^2}v_iv_k\rg)  \xi_9
$$

              \ses  \ses

$$
-2(N-1)(b\wt k)b_ib_k+ 2(N-2)\xi_9   r_{ik} - 2(b\wt k) r_{ik}
+ \fr{gbq}{B} (b\wt k)  r_{ik}
$$

\ses\ses

$$
+\fr {gq}{B}\Biggl((N-1)(b\wt k)bb_k-(N-2)\xi_9   v_k +(b\wt k)\fr{q^2}b b_k  +(b\wt k)\fr{q^2}b e_k  \Biggr)
e_i
$$

\ses\ses

$$
+  (b\wt k)\fr {gq}{b}b_ke_i
- \fr {gq^3}{Bb}
\Bigl[b_i b_k + b_i e_k   +b_ke_i\Bigr]
 (b\wt k)
 $$

\ses

\ses\ses

$$
- N\fr {gq}b  (b \wt k) b_k\lf(b_i+\fr{q^2}Be_i  -\fr{q^2}Be_i\rg)
$$

\ses\ses

\be
-\fr {g^2}{b}  (b\wt k)  v_kb_i  -\fr {g^2}{b}  (b\wt k)  v_ke_i
+  \fr {g^3q}{B}
(b\wt k) e_i  v_k,
\ee

{

\nin
or
$$
R^m{}_{ikm}         +   R_i{}^h{}_{hk}
=
\fr N2 \fr {g^2q^2}{B}   (b\wt k)b_k   e_i
$$

\ses

\ses

$$
+\fr12  \fr {g^2q^2}{B}  (b \wt k)  \lf( r_{ik}-\fr1{q^2}v_iv_k\rg)
+ \fr {gq}{B}(b+gq)    \lf( r_{ik}-\fr1{q^2}v_iv_k\rg)  \xi_9
$$

\ses
\ses

$$
+ \Bigl[2(N-2)\xi_9   - 2(b\wt k) + \fr {gbq}{B} (b\wt k) \Bigr] \lf( r_{ik}-\fr1{q^2}v_iv_k\rg)
$$

\ses

\ses

$$
+ \fr1b
\Bigl[2(N-2)\xi_9  - 2(b\wt k)   \Bigr]
v_k(b_i+e_i)
+
 \fr {gq}{B} (b\wt k)
v_k   (e_i+b_i)
$$

              \ses  \ses

$$
-2(N-1)(b\wt k)b_k        \lf(b_i+\fr{q^2}Be_i - \fr{q^2}Be_i\rg)
-(N-2)\fr {gq}{B}\xi_9   v_k   e_i
$$

\ses\ses

$$
-\fr {gbq}{B}(b\wt k)b_ke_i
+\fr {gq^3}{Bb}(b\wt k) e_k e_i
+  (b\wt k)\fr {gq}{b}b_ke_i
 $$

\ses\ses

$$
- \fr {gq^3}{Bb}
 (b\wt k)\Bigl[b_k + e_k  \Bigr]
b_i
- N\fr {gq}b  (b \wt k) b_k\lf(b_i+\fr{q^2}Be_i\rg)
$$

\ses\ses

$$
-\fr {g^2}{b}  (b\wt k)  v_k\lf(b_i+\fr{q^2}Be_i - \fr{q^2}Be_i\rg)  -  \fr {g^2}{Bb}     (b^2+q^2)
(b\wt k) e_i  v_k.
$$

{

The last expression can conveniently written as
$$
R^m{}_{ikm}         +   R_i{}^h{}_{hk}
=
\fr N2 \fr {g^2q^2}{B}   (b\wt k)b_k   e_i
$$

\ses

\ses

$$
+\fr12  \fr {g^2q^2}{B}  (b \wt k)  \lf( r_{ik}-\fr1{q^2}v_iv_k\rg)
+ \fr {gq}{B}(b+gq)    \lf( r_{ik}-\fr1{q^2}v_iv_k\rg)  \xi_9
$$

\ses
\ses

$$
+ \Bigl[2(N-2)\xi_9   - 2(b\wt k) + \fr {gbq}{B}   (b\wt k) \Bigr] \lf( r_{ik}-\fr1{q^2}v_iv_k\rg)
$$

\ses
\ses

$$
+\fr1b
 \Bigl[2(N-2)\xi_9  - 2(b\wt k)   \Bigr]
v_k
\lf[e_i+b_i+\fr{q^2}Be_i - \fr{q^2}Be_i\rg]
+
 \fr {gq}{B} (b\wt k)
v_k e_i
$$

              \ses  \ses

$$
-2(N-1)(b\wt k)b_k        \lf(b_i+\fr{q^2}Be_i - \fr{q^2}Be_i\rg)
-(N-2)\fr {gq}{B}\xi_9   v_k   e_i
$$

\ses\ses

$$
+\fr {gq^3}{Bb}(b\wt k) e_k e_i
+  \fr{gq^2}{bB}(q+gb)(b\wt k)b_ke_i
 $$

\ses\ses

\be
- N\fr {gq}b  (b \wt k)\lf(b_i+\fr{q^2}Be_i\rg)
 b_k
-\fr {g^2}{b}  (b\wt k) \lf(b_i+\fr{q^2}Be_i \rg)
 v_k
-  \fr {g^2b}{B}
(b\wt k) e_i  v_k.
\ee

{

Next,  we use            the equality
\be
y_i=(q^2e_i+Bb_i)  \fr{K^2}{bB}
\ee
(see  (A.23) in Appendix A and (C.45) in Appendix C),
obtaining
\be
R^m{}_{ikm}         +   R_i{}^h{}_{hk}
=
2X   \lf( r_{ik}-\fr1{q^2}v_iv_k\rg)
+2Y_ke_i+\fr 2{K^2}Z_ky_i
\ee
\ses
with
\be
X=
(N-2)\xi_9   - (b\wt k) + \fr12 \fr {gbq}{B}  (b\wt k)
+\fr14  \fr {g^2q^2}{B}  (b \wt k)
+  \fr12 \fr {gq}{B}(b+gq)     \xi_9
\ee
\ses
and
$$
Z_k=
 \Bigl[(N-2)\xi_9  - (b\wt k)\Bigr]
v_k
-(N-1) b (b\wt k)
b_k
 - \fr N2 gq
(b\wt k) b_k
  -  \fr12 g^2  (b\wt k)
 v_k.
$$

{

Insert here $v_k=(q^2/b)(e_k+b_k)$:
$$
Z_k=
 \Bigl[(N-2)\xi_9  - (b\wt k)\Bigr]\fr{q^2}b e_k  +
 \Biggl[   \Bigl[(N-2)\xi_9  - (b\wt k)\Bigr]\fr{q^2}b
-(N-1) b (b\wt k)
\Biggr]
b_k
$$
\ses
$$
 - \fr N2 gq
(b\wt k) b_k
  -  \fr12 g^2  (b\wt k)\fr{q^2}b e_k       -  \fr12 g^2  (b\wt k)\fr{q^2}b b_k.
$$
Apply (D.4):
$$
Z_k=
 \Bigl[(N-2)\xi_9  - (b\wt k)\Bigr]\fr{q^2}b e_k  +
 \Biggl[   \Bigl[(N-2)  \xi_9  - (b\wt k)\Bigr]q^2
-(N-1) b^2 (b\wt k)
\Biggr]
\fr 1{K^2}y_k
$$
\ses
$$
- \Biggl[   \Bigl[(N-2)\xi_9  - (b\wt k)\Bigr]q^2
-(N-1) b^2 (b\wt k)
\Biggr]\fr{q^2}{bB}e_k
$$
\ses
$$
 - \fr N2 gq
(b\wt k)\Bigl(\fr b{K^2}y_k-\fr{q^2}Be_k\Bigr)
  -  \fr12 g^2  (b\wt k)\fr{q^2}b e_k       -  \fr12 g^2  (b\wt k)\fr{q^2}b \Bigl(\fr b{K^2}y_k-\fr{q^2}Be_k\Bigr),
$$
\ses
so that
\be
Z_k=Z_{\{y\}}\fr1{K^2} y_k+Z_{\{e\}}\fr{q^2}B  e_k
\ee
\ses
with
\be
Z_{\{y\}}=
   \Bigl[(N-2)  \xi_9  - (b\wt k)\Bigr]q^2
-(N-1) b^2 (b\wt k)
 - \fr N2 gbq
(b\wt k)
     -  \fr12 g^2  (b\wt k)q^2
\ee
and
$$
Z_{\{e\}}=
 \Bigl[(N-2)\xi_9  - (b\wt k)\Bigr]\fr{B}b
- \Biggl[   \Bigl[(N-2)\xi_9  - (b\wt k)\Bigr]q^2
-(N-1) b^2 (b\wt k)
\Biggr]\fr{1}{b}
$$
\ses
$$
+ \fr N2 gq
(b\wt k)
  -  \fr12 g^2  (b\wt k)\fr{B}b      +  \fr12 g^2  (b\wt k)\fr{q^2}b,
$$
which can be written as
\be
Z_{\{e\}}=
(N-2)\xi_9 \fr{B-q^2}{b}
+ (N-2)b (b\wt k)
+ \fr12(N-2) gq
(b\wt k)
  -  \fr12 g^2  (b\wt k)\fr{B}b      +  \fr12 g^2  (b\wt k)\fr{q^2}b.
\ee

{

Also, we get
$$
Y_k=      \fr N 4 \fr {g^2q^2}{B}   (b\wt k)b_k
+ (N-2)\fr{b+gq}B   \xi_9   v_k
-\fr{b}B  (b\wt k)   v_k
-\fr{gq}B  (b\wt k)   v_k
$$

\ses

$$
+\fr12\fr{gq}B(b\wt k)v_k
+(N-1)(b\wt k) \fr{q^2}Bb_k
$$

\ses

$$
-  \fr12(N-2)\fr {gq}{B}\xi_9   v_k
+\fr {gq^3}{2Bb}(b\wt k) e_k
+  \fr12 \fr{gq^3}{bB}(b\wt k) b_k
-  \fr12  \fr {g^2q^2}{B}
(b\wt k)  e_k,
$$

\ses

or
$$
Y_k=      \fr N 4 \fr {g^2q^2}{B}   (b\wt k)b_k
+ (N-2)\fr{b+gq}B   \xi_9   v_k
-\fr{b}B  (b\wt k)   v_k
$$

\ses

$$
+(N-1)(b\wt k) \fr{q^2}Bb_k
-  \fr12(N-2)\fr {gq}{B}\xi_9   v_k
-  \fr12  \fr {g^2q^2}{B} (b\wt k)  e_k.
$$
Inserting  here $v_k=(q^2/b)(e_k+b_k)$ yields
$$
Y_k=
 \fr N 4 \fr {g^2q^2}{B}   (b\wt k)b_k
+ (N-2)\fr{b+gq}B   \xi_9   \fr{q^2}b e_k  +  (N-2)\fr{b+gq}B   \xi_9   \fr{q^2}b b_k
-\fr{q^2}B  (b\wt k)  e_k   -\fr{q^2}B  (b\wt k)   b_k
$$

\ses

$$
+(N-1)(b\wt k) \fr{q^2}Bb_k
-  \fr12(N-2)\fr {gq}{B}\xi_9     \fr{q^2}b e_k   -    \fr12(N-2)\fr {gq}{B}\xi_9     \fr{q^2}b b_k
-  \fr12  \fr {g^2q^2}{B} (b\wt k)  e_k.
$$

\ses

{

or
$$
Y_k=
 \fr N 4 \fr {g^2q^2}{B}   (b\wt k)b_k
+ (N-2)\fr{q^2}B   \xi_9  e_k  +  (N-2)\fr{q^2}B   \xi_9   b_k
-\fr{q^2}B  (b\wt k)  e_k
$$

\ses\ses

$$
+(N-2)(b\wt k) \fr{q^2}Bb_k
+  \fr12(N-2)\fr {gq}{B}\xi_9     \fr{q^2}b e_k   +    \fr12(N-2)\fr {gq}{B}\xi_9     \fr{q^2}b b_k
-  \fr12  \fr {g^2q^2}{B} (b\wt k)  e_k,
$$
so that,
\be
Y_k=\fr{q^2}B      Y_{\{y\}}\fr1{K^2} y_k+\fr{q^2}B     Y_{\{e\}}e_k
\ee
with
\be
Y_{\{y\}}=
  (N-2)b \xi_9
+n_7  (b\wt k) b
+ \fr12(N-2)   gq\xi_9
\ee
and
$$
Y_{\{e\}}=
 \Bigl((N-2)\xi_9-(b\wt k)\Bigr)  -  (N-2)\fr{q^2}B   \xi_9
$$

\ses\ses

$$
-n_7(b\wt k) \fr{q^2}B
+  \fr12(N-2)\fr {gq}{b}\xi_9
-    \fr12(N-2)\fr {gq}{B}\xi_9     \fr{q^2}b
-  \fr12  g^2 (b\wt k),
$$
\ses
or
\be
Y_{\{e\}}=
(N-2)\xi_9-(b\wt k)  -  (N-2)\fr{q^2}B   \xi_9
-n_7(b\wt k) \fr{q^2}B
+  \fr12(N-2)  \fr{gq}  B  (b+gq)   \xi_9
-  \fr12  g^2 (b\wt k),
\ee
where we have introduced the notation
\be
n_7=N-2+\fr N4g^2.
\ee

Since
$y^ke_k=0$, from (D.7) and (D.10) it ensues that
\be
Z_ky^k=  Z_{\{y\}},   \qquad   Y_ky^k=   \fr{q^2}B      Y_{\{y\}}.
\ee

The function  (D.11) obeys the equality
\be
Y_{\{y\}}=s_1,
\ee
where the right--hand part is the function (C.60) of Appendix C.

{

We may use the angular metric tensor
\be
h_{ij}=\fr1{A_lA^l}A_iA_j+     \lf( r_{ij}-\fr1{q^2}v_iv_j\rg)    \fr{K^2}B
\ee
and the contracted Cartan tensor
\ses
\be
A_i=
-\fr {NK}2g\fr q{B}e_i,  \qquad   A_hA^h=\fr{N^2}{4}g^2,
\ee
obtaining
$$
R^m{}_{ikm}         +   R_i{}^h{}_{hk}
=
2X  \fr B{K^2} \lf[ h_{ki}-\fr1{A_lA^l}A_iA_k\rg]
-\fr{2B}{NKgq} 2Y_kA_i+\fr2{K^2}Z_ky_i.
$$

\ses

With this result, we obtain  the tensor
$\rho^i{}_k$
to read explicitly as
\ses
\be
\rho^i{}_k=
X  \fr B{K^2} \lf[ h_k^i-\fr1{A_lA^l}A^iA_k\rg]
-\fr{ 2B}{NKgq}Y_kA^i+\fr1{K^2}Z_ky^i
-\fr12\de_k^i  R^{nh}{}_{hn}.
\ee
\ses
Here,
\be
h_k^i=\de_k^i-\fr1{K^2}    y^iy_k,
\ee
\ses
so that
$$
\rho^i{}_k=
\fr 1{2K^2}M_7\lf[ h_k^i-\fr1{A_lA^l}A^iA_k\rg]
-\fr12\ \fr1{A_lA^l}A^iA_k     R^{nh}{}_{hn}
$$

\ses\ses

\be
-\fr{ 2B}{NKgq}Y_kA^i+\fr1{K^2}Z_ky^i
-\fr12 \fr1{K^2}    y^iy_k     R^{nh}{}_{hn},
\ee
\ses
where
\be
M_7=2BX- K^2 R^{nh}{}_{hn}.
\ee

{

From (C.48) we know that
\ses
$$
K^2 R^{ni}{}_{in}
=
- 2 (N-1)(b\wt k)B
 +  (N-1) (N-2) B \xi_9
$$

\ses

\be
+\fr12 (N-2) g^2q^2 (b\wt k)
+  (N-2) gq(b+gq)\xi_9
+  (N-2) gbq  (b\wt k).
\ee
\ses
By the help of (D.6) and (D.22), the scalar (D.21) reduces to
$$
M_7
=
 2 (N-2)(b\wt k)B
-  (N-3) (N-2) B \xi_9
$$

\ses

$$
-\fr12 (N-3) g^2q^2 (b\wt k)
-  (N-3) gq(b+gq)\xi_9
-  (N-3) gbq  (b\wt k).
$$
In terms of the function $\cP$ (see (C.68) in Appendix C)
this scalar can be written as
$$
M_7
=  2B\cP-\fr N2g^2B(b\wt k)
$$

\ses

\be
-\fr12 (N-3) g^2q^2 (b\wt k)
-  (N-3) gq(b+gq)\xi_9
-  (N-3) gbq  (b\wt k).
\ee

{

Using (D.7) and (D.11) in (D.20) leads to the expansion
$$
\rho^i{}_k=
\fr 1{2K^2}M_7\lf[ h_k^i-\fr1{A_lA^l}A^iA_k\rg]
-\fr12\ R^{nh}{}_{hn} \fr1{A_lA^l}A^iA_k
+\fr B{K^2}Y_{\{e\}} \fr1{A_lA^l}A^iA_k
$$

\ses

$$
  -\fr{ 2q}{NKg} \fr 1{K^2}     \Bigl(Y_{\{y\}}y_kA^i + Z_{\{e\}}A_ky^i \Bigr)
 +\fr1{K^2}  \fr1{K^2}   Z_{\{y\}}y^iy_k
-\fr12 \fr1{K^2}   R^{nh}{}_{hn}  y^iy_k    ,
$$
\ses
or
\be
\rho^i{}_k= \fr 1{2K^2}M_7 h_k^i
+\fr 1{2K^2}M_8\fr1{A_lA^l}A^iA_k
  -\fr{ 2q}{NKg}\fr 1{K^2}   \Bigl(Y_{\{y\}}y_kA^i + Z_{\{e\}}A_ky^i \Bigr)
+\fr 1{2K^2}M_9   \fr1{K^2}  y^iy_k,
\ee
where
\be
M_8=-M_7 - K^2 R^{nh}{}_{hn} +2BY_{\{e\}}=-2BX +2BY_{\{e\}}
\ee
and
\be
M_9=
2  Z_{\{y\}}
-K^2   R^{nh}{}_{hn}.
\ee

{

Simple direct calculations yield
\be
M_8=
 - gbq  (b\wt k)
-\fr12 g^2q^2  (b \wt k)
  -  2(N-2)q^2  \xi_9
-2 n_7   (b\wt k)q^2
+  (N-3)gq    (b+gq)\xi_9
-   g^2B (b\wt k)
\ee

{

and

$$
M_9=
2    \Bigl[(N-2)  \xi_9  - (b\wt k)\Bigr]q^2
-2(N-1) b^2 (b\wt k)
 - N gq
(b\wt k) b
     -   g^2  (b\wt k)q^2
$$
\ses

$$
+ 2 (N-1)(b\wt k)B
-  (N-1) (N-2) B \xi_9
$$

\ses

$$
-\fr12 (N-2) g^2q^2 (b\wt k)
-  (N-2) gq(b+gq)\xi_9
-  (N-2) gbq  (b\wt k).
$$

\ses\ses

$$
=
2(N-2)  \xi_9 q^2
-2(N-2) b^2 (b\wt k)
 - (N-2) gbq
(b\wt k)
$$
\ses

$$
+ 2 (N-2)(b\wt k)(b^2+gbq+q^2)
-  (N-1) (N-2) B \xi_9
$$

\ses

$$
-\fr12 N g^2q^2 (b\wt k)
-  (N-2) gq(b+gq)\xi_9
-  (N-2) gbq  (b\wt k),
$$
\ses
or
$$
M_9=
2(N-2)  \xi_9 q^2
-  (N-1) (N-2) B \xi_9
+2\lf(N-2-\fr N4g^2\rg)   q^2 (b\wt k)
$$
\ses
\be
-  (N-2) gq(b+gq)\xi_9.
\ee

{

Now we get
\be
\rho^i{}_k= \fr 1{2K^2}M_7    \de_k^i
+\fr 1{2K^2}M_8\fr1{A_lA^l}A^iA_k
  -\fr{ 2q}{NKg}\Bigl(Y_{\{y\}}y_kA^i + Z_{\{e\}}A_ky^i \Bigr)
+\fr 1{2K^2}M_{10}   \fr1{K^2}  y^iy_k,
\ee
where
$$
M_{10}=M_9 - M_7.
$$
\ses
We have
$$
M_{10}=
2(N-2)  \xi_9 q^2
-  (N-1) (N-2) B \xi_9
+2\lf(N-2-\fr N4g^2\rg)   q^2 (b\wt k)
-  (N-2) gq(b+gq)\xi_9
$$


$$
- 2 (N-2)(b\wt k)B
+  (N-3) (N-2) B \xi_9
$$

\ses

$$
+\fr12 (N-3) g^2q^2 (b\wt k)
+  (N-3) gq(b+gq)\xi_9
+  (N-3) gbq  (b\wt k),
$$
\ses
or
$$
M_{10}=
- 2 (N-2)(b\wt k)(B-q^2)
-2 (N-2) (B-q^2) \xi_9
$$

\ses

\be
-\fr32 g^2q^2 (b\wt k)
- gq(b+gq)\xi_9
+  (N-3) gbq  (b\wt k).
\ee

{

Now we may write:
\be
\rho^{ik}= \fr 1{2K^2}M_7  g^{ik}
+\fr 1{2K^2}M_8\fr1{A_lA^l}A^iA^k
  -\fr{ 2q}{NKg}\fr 1{K^2}       \Bigl(Y_{\{y\}}y^kA^i + Z_{\{e\}}A^ky^i \Bigr)
+\fr 1{2K^4}M_{10}  y^iy^k.
\ee

Using in (D.31) the representations
\be
g^{ik}=
\biggl[a^{ik}+\fr gq(bb^ib^k-b^iy^k-b^ky^i)+\fr g{Bq}(b+gq)y^iy^k
\biggr]\fr B{K^2}
\ee
and
\be
A^i=\fr N2g\fr 1{qK}
\Bigl[Bb^i-(b+gq)y^i\Bigr],
\ee
 we arrive at the expansion
\be
\rho^{ik}= E_1 a^{ik}
+E_2b^ib^k+E_{3Y}b^iy^k+E_{3Z}b^ky^i + E_4y^iy^k,
\ee
where
\be
E_1=\fr B{2K^4}M_7,
\ee

\ses

\be
E_2=    \fr B{2K^4q^2}(gbq M_7   +BM_8),
\ee

\ses

\be
E_{3Y}=-\fr{B}{2K^4}
\Bigl(\fr g{q}M_7
+\fr1{q^2}(b+gq)M_8+ 2 Y_{\{y\}}  \Bigr),
\ee

\ses

\be
E_{3Z}=-\fr{B}{2K^4}
\Bigl(\fr g{q}M_7
+\fr1{q^2}(b+gq)M_8+ 2 Z_{\{e\}}  \Bigr),
\ee

\ses

\be
E_4=  \fr{1}{2K^4}\Biggl[\fr g{q}(b+gq)M_7+
\fr1{q^2}(b+gq)^2M_8
+    2(b+gq)\Bigl( Y_{\{y\}}+ Z_{\{e\}} \Bigr)
+
M_{10}
\Biggr].
\ee

{

\ses \ses

Let us perform the required calculation:
$$
-\fr2{B} K^4E_{3Y}= \fr g{q}
\Biggl(
2B\cP-\fr N2g^2B(b\wt k)
$$

\ses

$$
-\fr12 (N-3) g^2q^2 (b\wt k)
-  (N-3) gq(b+gq)\xi_9
-  (N-3) gbq  (b\wt k)
\Biggr)
$$

\ses\ses

$$
+\fr b{q^2}
\Biggl[
 - gbq  (b\wt k)
-\fr12 g^2q^2  (b \wt k)
- gq(b+gq)     \xi_9
 -  2(N-2)q^2  \xi_9
$$

\ses\ses

$$
-2 n_7   (b\wt k)q^2
+  (N-2)gq(b+gq)   \xi_9
-   g^2B (b\wt k)
\Biggr]
$$

\ses

\ses\ses

$$
+ g    \fr1{q}
\Biggl[
 - gbq  (b\wt k)
-\fr12 g^2q^2  (b \wt k)
- gq(b+gq)     \xi_9
 -  2(N-2)q^2  \xi_9
$$

\ses\ses

$$
-2 n_7   (b\wt k)q^2
+  (N-2) gq(b+gq)    \xi_9
-   g^2B (b\wt k)
\Biggr]
$$

\ses

\ses\ses

$$
+
2  (N-2)b \xi_9
+2n_7  (b\wt k) b
+ (N-2)   gq\xi_9.
$$
\ses
Here, all the terms which are proportional to  $g$ are cancelled, leaving us with
\be
E_{3Y}=gT_{3Y},
\ee

{

where
$$
-\fr 2{B} K^4T_{3Y}= \fr 1{q}
\Biggl(
2B\cP-\fr N2g^2B(b\wt k)
-\fr12 (N-2) g^2q^2 (b\wt k)
-  (N-2) gbq  (b\wt k)
\Biggr)
$$

\ses\ses

$$
+\fr b{q}
\Biggl[
 - b  (b\wt k)
-\fr12  gq  (b \wt k)
- (b+gq)     \xi_9
+  (N-2)(b+gq) \xi_9
                                     -  \fr1q   gB (b\wt k)
\Biggr]
$$

\ses\ses

\be
+    \fr1{q}
\Biggl[
 -  (N-2)q^2  \xi_9
-2\lf(N-2+\fr N4g^2\rg)   (b\wt k)q^2
-   g^2B (b\wt k)
\Biggr].
\ee

{


Also,
$$
-\fr2{B} K^4E_{3Z}= \fr g{q}
\Biggl(
2B\cP-\fr N2g^2B(b\wt k)
$$

\ses

$$
-\fr12 (N-3) g^2q^2 (b\wt k)
-  (N-3) gq(b+gq)\xi_9
-  (N-3) gbq  (b\wt k)
\Biggr)
$$

\ses\ses

$$
+\fr b{q^2}
\Biggl[
 - gbq  (b\wt k)
-\fr12 g^2q^2  (b \wt k)
- gq(b+gq)     \xi_9
 -  2(N-2)q^2  \xi_9
$$

\ses\ses

$$
-2\lf(N-2+\fr N4g^2\rg)   (b\wt k)q^2
+  (N-2)gq(b+gq)   \xi_9
-   g^2B (b\wt k)
\Biggr]
$$

\ses

\ses\ses

$$
+ g    \fr1{q}
\Biggl[
 - gbq  (b\wt k)
-\fr12 g^2q^2  (b \wt k)
- gq(b+gq)     \xi_9
 -  2(N-2)q^2  \xi_9
$$

\ses\ses

$$
-2 n_7   (b\wt k)q^2
+  (N-2) gq(b+gq)    \xi_9
-   g^2B (b\wt k)
\Biggr]
$$

\ses

\ses\ses

$$
+
2(N-2)\xi_9 (b+gq)
+2 (N-2)b (b\wt k)
+ (N-2) gq
(b\wt k)
  - g^2  (b\wt k)(b+gq),
$$
\ses
so that
\be
E_{3Z}=gT_{3Z}
\ee

{

\nin
with
$$
-\fr 2{B} K^4T_{3Z}= \fr 1{q}
\Biggl(
2B\cP-\fr N2g^2B(b\wt k)
-\fr12 (N-2) g^2q^2 (b\wt k)
-  (N-2) gbq  (b\wt k)
\Biggr)
$$

\ses\ses

$$
+\fr b{q}
\Biggl[
 - b  (b\wt k)
-\fr12 gq  (b \wt k)
- (b+gq)     \xi_9
-       \fr N2 gq(b\wt k)
+  (N-2)(b+gq) \xi_9
                                     -  \fr1q   gB (b\wt k)
\Biggr]
$$

\ses\ses

\be
-    \fr1{q}
\Biggl[
2\lf(N-2+\fr N4g^2\rg)   (b\wt k)q^2
+   g^2B (b\wt k)
\Biggr]
+  (N-2) q
(b\wt k)
- g  (b\wt k)(b+gq).
\ee

\ses

We can infer that
\be
\fr 2{B} K^4(T_{3Y} -T_{3Z})=  T_7,
\ee
where $T_7$ is just the function (E.2) of Appendix E.

{

We continue:
$$
2K^4E_4= \fr g{q}(b+gq)
\Biggl(
 2 (N-2)(b\wt k)B
-  (N-3) (N-2) B \xi_9
$$

\ses\ses

$$
-\fr12 (N-3) g^2q^2 (b\wt k)
-  (N-3) gq(b+gq)\xi_9
-  (N-3) gbq  (b\wt k)
\Biggr)
$$

\ses\ses

$$
+\fr1{q^2}(b+gq)^2
\Biggl[
 - gbq  (b\wt k)
-\fr12 g^2q^2  (b \wt k)
- gq(b+gq)     \xi_9
 -  2(N-2)q^2  \xi_9
$$

\ses\ses

$$
-2 n_7   (b\wt k)q^2
+  (N-2)gq (b+gq)\xi_9
-   g^2B (b\wt k)
\Biggr]
$$

\ses\ses

$$
+              2(b+gq)
\Biggl[
  (N-2)b \xi_9
+n_7  (b\wt k) b
+ \fr12(N-2)   gq\xi_9
\Biggr]
$$

\ses\ses

$$
+           2(b+gq)
\Biggl[
(N-2)\xi_9 (b+gq)
+ (N-2)b (b\wt k)
+ \fr12(N-2) gq
(b\wt k)
  -  \fr12 g^2  (b\wt k)(b+gq)
  \Biggr]
$$

\ses\ses

$$
- 2 (N-2)(b\wt k)(B-q^2)
-2 (N-2) (B-q^2) \xi_9
$$

\ses\ses

$$
-\fr32 g^2q^2 (b\wt k)
- gq(b+gq)\xi_9
+  (N-3) gbq  (b\wt k),
$$
so that
\be
E_4=gT_4
\ee

{

and
$$
2K^4T_4=   \fr 2{q}(b+gq)B\cP
$$

\ses

$$
-\fr b{q}
\Biggl(
\fr N2g^2B(b\wt k)
+\fr12 (N-3) g^2q^2 (b\wt k)
+  (N-3) gq(b+gq)\xi_9
+  (N-3) gbq  (b\wt k)
\Biggr)
$$

\ses\ses

$$
- g
\Biggl(
\fr N2g^2B(b\wt k)
+\fr12 (N-3) g^2q^2 (b\wt k)
+  (N-3) gq(b+gq)\xi_9
+  (N-3) gbq  (b\wt k)
\Biggr)
$$

                        \ses\ses

\ses\ses

$$
+\fr{ b^2}{q^2}
\Biggl[
 - bq  (b\wt k)
-\fr32 gq^2  (b \wt k)
- q(b+gq)     \xi_9
+  (N-2)q (b+gq)   \xi_9
-   gB (b\wt k)
\Biggr]
$$

\ses


\ses\ses

$$
+2\fr b{q}
\Biggl[
 - gbq  (b\wt k)
-\fr12 g^2q^2  (b \wt k)
- gq(b+gq)     \xi_9
 -  2(N-2)q^2  \xi_9
$$

\ses\ses

$$
-2 n_7   (b\wt k)q^2
+  (N-2) gq (b+gq)  \xi_9
-   g^2B (b\wt k)
\Biggr]
$$
                                     \ses\ses

\ses


\ses

$$
+
g\Biggl[
 - gbq  (b\wt k)
-\fr12 g^2q^2  (b \wt k)
- gq(b+gq)     \xi_9
 -  2(N-2)q^2  \xi_9
$$

\ses\ses

$$
-2 n_7   (b\wt k)q^2
+  (N-2) gq  (b+gq)  \xi_9
-   g^2B (b\wt k)
\Biggr]
$$

\ses

\ses

$$
+             (N-2)   bq\xi_9
+              2q \Biggl[   (N-2)b \xi_9 +n_7  (b\wt k) b + \fr12(N-2)   gq\xi_9 \Biggr]
$$

\ses

$$
+
2(N-2)bq\xi_9
+ (N-2)bq
(b\wt k)
+           2q
\Biggl[
(N-2)gq\xi_9
+ \fr12(N-2) gq
(b\wt k)
  \Biggr]
$$

\ses\ses

$$
-\fr 32 gq^2 (b\wt k)
- q(b+gq)\xi_9
+  (N-3) bq  (b\wt k).
$$

{

We can simplify as follows:
$$
2K^4T_4=   \fr 2{q}(b+gq)B\cP
$$

\ses

$$
-\fr b{q}
\Biggl(
\fr N2g^2B(b\wt k)
+\fr12 (N-3) g^2q^2 (b\wt k)
+  (N-3) gq(b+gq)\xi_9
+  (N-3) gbq  (b\wt k)
\Biggr)
$$

\ses\ses

$$
- g
\Biggl(
\fr N2g^2B(b\wt k)
+\fr12 (N-2) g^2q^2 (b\wt k)
+  (N-3) gbq  (b\wt k)
\Biggr)
$$

\ses\ses

$$
+\fr{ b^2}{q^2}
\Biggl[
 - bq  (b\wt k)
-\fr32 gq^2  (b \wt k)
+  (N-3)q (b+gq)   \xi_9
-   gB (b\wt k)
\Biggr]
$$


\ses\ses

$$
+\fr b{q}
\Biggl[
 - 2gbq  (b\wt k)
-3 g^2q^2  (b \wt k)
+  2(N-3) gq (b+gq)  \xi_9
-   2g^2B (b\wt k)
\Biggr]
$$
                                     \ses\ses


$$
+
g\Biggl[
-2\lf(N-2+\fr N4g^2\rg)   (b\wt k)q^2
-   g^2B (b\wt k)
\Biggr]
$$

\ses

\ses


$$
+             (N-2)   bq\xi_9
+              (N-2)   gq^2\xi_9
+ (N-2)bq
(b\wt k)
+
(N-2) gq^2
(b\wt k)
$$

\ses\ses

\be
-\fr 32 gq^2 (b\wt k)
- q(b+gq)\xi_9
+  (N-3) bq  (b\wt k).
\ee

{

In (D.20)
 we may apply the formula
$$
A^i=\fr N2g\fr 1{qK}
\Bigl[Bb^i-(b+gq)y^i\Bigr]
$$
(see (D.33)) and obtain
\be
\rho^i{}_k=
\fr 1{2K^2}M_7\lf[ h_k^i-\fr1{A_lA^l}A^iA_k\rg]
-\fr12\ \fr1{A_lA^l}A^iA_k     R^{nh}{}_{hn}
-\fr{ B^2}{K^2q^2}Y_k  b^i
+\fr1{K^2}    P_ky^i,
\ee
\ses
where
\be
P_k=\fr{ B}{q^2}(b+gq)  Y_k
+Z_k
-\fr12 y_k     R^{nh}{}_{hn}.
\ee
We can decompose this vector:
\be
P_k=P_{\{y\}}\fr1{K^2} y_k+P_{\{e\}}\fr{q^2}B  e_k
\ee
and use (D.7)--(D.12), getting
$$
P_{\{y\}}=
(b+gq)Y_{\{y\}} + Z_{\{y\}}   -\fr12   K^2     R^{nh}{}_{hn}
$$

\ses

$$
=
b
\Biggl(  (N-2)b \xi_9
+n_7  (b\wt k) b
+ \fr12(N-2)   gq\xi_9
\Biggr)
$$

\ses

$$
+gq
\Biggl(  (N-2)b \xi_9
+n_7  (b\wt k) b
+ \fr12(N-2)   gq\xi_9
\Biggr)
$$

\ses

$$
+   \Bigl[(N-2)  \xi_9  - (b\wt k)\Bigr]q^2
-(N-1) b^2 (b\wt k)
 - \fr N2 gbq
(b\wt k)
     -  \fr12 g^2  (b\wt k)q^2
$$

\ses

$$
+ (N-1)(b\wt k)B
 -\fr12  (N-1) (N-2) B \xi_9
$$

\ses

$$
-\fr14 (N-2) g^2q^2 (b\wt k)
-\fr12  (N-2) gq(b+gq)\xi_9
-\fr12  (N-2) gbq  (b\wt k).
$$

{

Here, several reductions are available, namely,
$$
P_{\{y\}}
=
n_7  (b\wt k) b^2
+n_7  (b\wt k) gbq
  - q^2(b\wt k)
-(N-1) b^2 (b\wt k)
 - \fr N2 gbq
(b\wt k)
$$

\ses

$$
+ (N-1)(b\wt k)B
 -\fr12  (N-3) (N-2) B \xi_9
-\fr14 N g^2q^2 (b\wt k)
-\fr12  (N-2) gbq  (b\wt k)
$$

            \ses\ses

$$
=
\lf(N-2+\fr N4g^2\rg) (b\wt k) b^2
+   \lf(N-2+\fr N4g^2\rg)  (b\wt k) gbq
$$

\ses

$$
-(N-2) b^2 (b\wt k)
(b\wt k)
+ (N-2)(b\wt k)B
 -\fr12  (N-3) (N-2) B \xi_9
-\fr14 N g^2q^2 (b\wt k)
-  (N-2) gbq  (b\wt k)
$$

\ses

$$
  =  (N-2)(b\wt k)B
 - \fr12 (N-2) (N-3) B \xi_9
+\fr14N  g^2   (b\wt k)(b^2+gbq)
-\fr14 N g^2q^2 (b\wt k).
$$
Eventually,
we arrive at the function
\be
P_{\{y\}}   =  n_7(b\wt k)B
 - \fr12 (N-2) (N-3) B \xi_9
-\fr12 N     g^2q^2 (b\wt k)
\ee
which is tantamount to the function (C.62) of Appendix C:
\be
P_{\{y\}}    =  P.
\ee


{

\setcounter{equation}{0}

\nin
{\bf\large Appendix E: ~  Skew part}

\ses\ses

 In the special case of Appendix C from (B.53) we can find the skew part
 $\rho_{ik}-\rho_{ki}$ of the covariantly conserved tensor $\rho_{ik}$:
$$
\rho_{ik}-\rho_{ki}=
\Biggl[(N-2)\fr g{B}q \fr 14 k^2g^2
- \fr12N \fr {g^2b}{B} (b\wt k) \Biggr]
(b_iv_k-b_kv_i)
$$

\ses

\ses

$$
+
\fr g{B}\Bigl (qb_i-b\fr{v_i}q\Bigr) y^j a_{njkm}  a^{nm}
-\fr g{B}\Bigl (qb_k-b\fr{v_k}q\Bigr) y^j a_{njim}  a^{nm}
$$

\ses\ses

$$
+ \fr g{Bq}(b+gq)y^my^n  ( a_{nikm}- a_{nkim})
$$

\ses

$$
 +(N-1)\fr g{Bq}  (q^2+b^2) (b\wt k)
 (b_iv_k-b_kv_i)
$$

\ses

$$
-\fr gq  b  \fr g{B}q (b\wt k) (b_iv_k-b_kv_i)
-\fr gq   \fr {g^2}{B} q^2(b\wt k) (b_iv_k-b_kv_i).
$$
\ses
Using  here the representations (C.42) and (C.43) specific for the isotropic case yields
$$
\rho_{ik}-\rho_{ki}=
\fr 14 (N-2)\fr g{B}q k^2g^2 (b_iv_k-b_kv_i)
-\fr12 N \fr {g^2}{B}b  (b\wt k)  (b_iv_k-b_kv_i)
$$

\ses

$$
-  \fr g{B}\Bigl (qb_i-b\fr{v_i}q\Bigr)   \Biggl(  (N-1)(b\wt k)bb_k-(N-2)\xi v_k +(b\wt k) v_k  \Biggr)
$$

\ses

$$
+  \fr g{B}\Bigl (qb_k-b\fr{v_k}q\Bigr)   \Biggl(  (N-1)(b\wt k)bb_i-(N-2)\xi v_i +(b\wt k) v_i  \Biggr)
$$

\ses

$$
+ (N-1)\fr g{Bq}  (q^2+b^2)  (b\wt k)(b_iv_k-b_kv_i)
$$

\ses

$$
-
\fr {g^2(b+gq)}{B}(b\wt k)(b_iv_k-b_kv_i),
$$

{

or
$$
\rho_{ik}-\rho_{ki}=
\fr 14 (N-2)\fr g{B}q k^2g^2 (b_iv_k-b_kv_i)
-\fr12 N \fr {g^2}{B}b  (b\wt k)  (b_iv_k-b_kv_i)
$$

\ses

$$
-  \fr {gq}{B}  \Biggl(  -(N-2)\xi  +(b\wt k)  \Biggr)(b_iv_k-b_kv_i)
- (N-1) \fr {gb^2}{qB}(b\wt k)   (b_iv_k-b_kv_i)
$$

\ses

$$
+ (N-1)\fr g{Bq}  (q^2+b^2)  (b\wt k)(b_iv_k-b_kv_i)
-
\fr {g^2(b+gq)}{B}(b\wt k)(b_iv_k-b_kv_i)
$$

\ses

$$
=
-\fr12 N \fr {g^2}{B}b  (b\wt k)  (b_iv_k-b_kv_i)
-  \fr {gq}{B}  \Biggl(  -(N-2)\xi_9  +(b\wt k)  \Biggr)(b_iv_k-b_kv_i)
$$

\ses

$$
+ (N-1)\fr {gq}{B} (b\wt k)(b_iv_k-b_kv_i)
-
\fr {g^2(b+gq)}{B}(b\wt k)(b_iv_k-b_kv_i).
$$

{

Eventually, we get
\be
\fr12(\rho_{ik}-\rho_{ki})=
 \fr {g}{2B} T_7 (b_iv_k-b_kv_i),
\ee
\ses
where
\be
T_7=
-\fr12 N gb  (b\wt k)
-g(b+gq)(b\wt k)
+(N-2) q \Bigl(\xi_9+ (b\wt k)  \Bigr).
\ee

If we  now apply the formulas
$$
y_i=\Bigl(v_i+(b+gq)b_i\Bigr)\fr{K^2}B, \qquad
A_i=\fr {NK}2g\fr1{q}(b_i-\fr b{K^2}y_i),
$$
we obtain
$$
\fr gB(b_iv_k-b_kv_i)=\fr g{K^2}(b_iy_k-b_ky_i)=
\fr g{K^2}\fr2{NK}\fr qg(A_iy_k-A_ky_i),
$$
so  that
\be
\fr gB(b_iv_k-b_kv_i) = \fr q{K^2}\fr2{N}(A_il_k-A_kl_i)
\ee
\ses
and we may  write (E.1) in the tensorial form
\be
\fr12(\rho_{ik}-\rho_{ki})=
\fr 1{NK^2}
q T_7 (A_il_k-A_kl_i).
\ee

From  (A.65) we can infer
\be
\rho_{ij}-\rho_{ji}   =    R_i{}^m{}_{mj}-R_j{}^m{}_{mi}  +R^m{}_{ijm}  -  R^m{}_{jim},
\ee
so that
\be
A^iy^j(R_i{}^m{}_{mj}-R_j{}^m{}_{mi}  +R^m{}_{ijm}  -  R^m{}_{jim})=\fr  4NqT_7A^iA_i,
\ee
or
\be
 Ng^2qT_7=A^iy^j(R_i{}^m{}_{mj}-R_j{}^m{}_{mi}  +R^m{}_{ijm}  -  R^m{}_{jim}).
\ee
This formula shows how the coefficient $T_7$ is expressed through the curvature tensor.

\ses

Since
$
(v_k)_{|_{y^h=b^h}}=0,
$
from (E.1) and (E.2) we should conclude that
\be
(\rho_{ik}-\rho_{ki}) _{|_{y^h=b^h}}=0.
\ee

\vskip 1cm

\def\bibit[#1]#2\par{\rm\noindent\parskip1pt
                     \parbox[t]{.05\textwidth}{\mbox{}\hfill[#1]}\hfill
                     \parbox[t]{.925\textwidth}{\baselineskip11pt#2}\par}

\nin
{\large  REFERENCES}

\ses

\bibit[1] S. Weinberg: \it Gravitation and Cosmology, \rm   John Wiley, New York 1972.

\ses

\bibit[2]  M. Kamionkowski:  Dark matter and dark energy, {\it  arXiv:astro-ph}/0706.2986 (2007).

\ses

\bibit[3] H. Rund: \it The Differential Geometry of Finsler
 Spaces, \rm Springer, Berlin 1959.

\ses

\bibit[4] G.S. Asanov: \it Finsler Geometry, Relativity and Gauge
 Theories, \rm D.~Reidel Publ. Comp., Dordrecht 1985.

\ses

 \bibit[5] D.~Bao, S.S. Chern, and Z. Shen: {\it  An
Introduction to Riemann-Finsler Geometry,}  Springer, N.Y., Berlin 2000.

\bibit[6] J.I. Horv\'ath: New geometrical methods of the theory of physical fields,
\it Nouv. Cim. \bf 9 \rm(1958), 444--496.

\bibit[7] Y. Takano: Theory of fields in Finsler spaces, \it Progr. Theor. Phys. \bf32 \rm(1968),  1159--1180.

\ses

\bibit[8] R. Mrugala: Riemannian and Finslerian geometry in thermodynamics,
\it Open Syst. Inform. Dyn. \bf1 \rm    (1992), 379--396.

\ses

\bibit[9] R.S. Ingarden and L. Tamassy:  On Parabolic geometry and irreversible macroscopic time,
 \it Rep. Math. Phys. \bf 32 \rm(1993), 11.

\ses

\bibit[10] R.S. Ingarden:  On physical applications of Finsler Geometry,
 \it Contemporary Mathematics \bf 196 \rm(1996), 213--223.

\ses

\bibit[11] G.S. Asanov:  Finsleroid--Finsler  space with Berwald and  Landsberg conditions,
 {\it  arXiv:math.DG}/0603472 (2006).

\ses

 \bibit[12] G.S. Asanov:  Finsleroid--Finsler  spaces of positive--definite and  relativistic types.
 \it Rep. Math. Phys. \bf 58 \rm(2006), 275--300.

\ses

\bibit[13]
 G.S. Asanov:   {\it Finsleroid--Finsler  parallelism,}  {\it  arXiv:math.DG}/0611345 (2006).

\ses

\bibit[14]
 G.S. Asanov:  Finsleroid--Finsler  parallelism. Cosmological aspects,
 \it Rep. Math. Phys. \bf 59 \rm(2007), 111--125.

\end{document}